# Salt and pepper noise removal method based on stationary Framelet transform with non-convex sparsity regularization

Yingpin Chen[1, 2]†, Yuming Huang[2], Lingzhi Wang[2], Huiying Huang[2], Jianhua Song[2], Chaoqun Yu[2], Yanping Xu[2]

(1 School of Mathematical Sciences, University of Electronic Science and Technology, 610054;

2 School of Physics and Information Engineering, Minnan Normal University, 363000)

**Abstract**: Salt and pepper noise occurs randomly and causes image degradation. Numerous denoising methods have been proposed to suppress this noise. However, existing methods have two main limitations. First, noise characteristics, such as noise location information and sparsity, are often described inaccurately or even ignored. Second, many existing methods separate the contaminated image into a recovered image and a noise part, leading to the recovery of an image with unsatisfactory smooth and detailed parts. In this study, we introduce a noise detection strategy to determine the position of the noise and employ a non-convex sparsity regularization depicted by $l_p$ quasi-norm to describe the sparsity of the noise, thereby addressing the first limitation. We adopt the morphological component analysis framework with stationary Framelet transform to decompose the processed image into the cartoon, texture, and noise parts to resolve the second limitation. Then, we apply the proposed model by using the alternating direction method of multipliers (ADMM). Finally, we conduct experiments to verify the proposed method and compare it with some current state-of-the-art denoising methods. The experimental results show that the proposed method can remove salt and pepper noise while preserving the details of the processed image and outperforming some state-of-the-art methods.

## 1. Introduction

Salt and pepper noise is a common impulse interference [1] in contaminated images, and it occurs during image acquisition, transmission, and decoding processes [2]. Salt noise is an impulse value that pollutes the pixel, and pepper noise is pixel contamination by a zero value [3]. Salt noise may be caused by random impulse interference, whereas pepper noise is mainly caused by a dysfunction of the image acquisition sensor. Information can be lost because of salt and pepper noise, resulting in interference of image analyses, such as object segmentation and object recognition. Therefore, salt and pepper denoising is essential in image processing. Fundamentally, the denoising process is an inverse problem that aims to obtain a reconstructed image from a polluted image [4-6].

† **Corresponding author**: Yingpin Chen is the corresponding author of this work. He is now an associate professor in Minnan Normal University. His E-mail is 110500617@163.com.

**Funding information**: This work is supported by the natural science foundation project of Fujian Province (2020J05169，2020J01816); the principal foundation of Minnan Normal University (KJ19019); the young and middle-aged teachers research and education project of the education department in Fujian Province (JAT190393, JAT190382, JAT210278); the high-level science research project of Minnan Normal University (GJ19019).

**Conflict of Interest statement**: This manuscript has not been published or presented elsewhere in part or in entirety and is not under consideration by another journal. We have read and understood your journal's policies, and we believe that neither the manuscript nor the study violates any of these. There are no conflicts of interest to declare.

**Permission to reproduce materials from other sources**: None.

**Data Availability statement:** All data, models, or code generated or used during the study are available from the corresponding author by request.

Common methods for removing salt and pepper noise include median-filter-based methods [7-16], total variation (TV)-based methods [17-21], Euler's elastica variational method [22], deep-learning-based methods [23-25], and wavelet transform (WT)-based methods [26]. Median-filter-based methods replace the noise pixel with the median value of adjacent pixels. These methods may distort the image structure and smoothness. TV-based methods have been widely used in image denoising. Although TV-based methods can preserve the edges of the processed image, they are prone to producing stair-case artifacts when processing the image smoothing area. Euler's elastica variational method lies in multiple parameters selection. Thus, the determination of optimal parameters for proper image recovery is challenging [22]. With the development of deep learning theory, various deep-learning-based models have been proposed for image denoising. But these methods require large amounts of sample data and high computational costs for training. WT-based methods regard the image as a sparse representation (SR) [27] of the wavelet dictionary. As salt and pepper noise occurs randomly, the wavelet cannot sparsely represent it. Thus, WT-based methods can reconstruct the image via a wavelet dictionary and remove the salt and pepper noise. However, the down-sampling operation of the WT-based methods may cause block artifacts in the recovery signal. Additionally, the WT only has one high-pass filter, which cannot accurately depict the detail of the processed signal. Researchers have made considerable efforts to improve the performance of WT-based methods. For instance, Wang *et al.* [28] proposed a non-down-sampling stationary WT to solve the block artifacts in the WT. Yan *et al.* [29] applied the Framelet transform (FT) [30, 31] to image denoising. Compared with the WT, the FT adds a high-pass filter analysis, which better depicts signal details.

Although the aforementioned methods have achieved good performance, these methods have two common limitations:

First, these methods do not fully explore the characteristics of salt and pepper noise. For example, salt-and-pepper noise location information is often ignored in traditional methods. Additionally, the detection of the noise location is easy because the amplitude characteristic of noise is unique. Furthermore, the sparsity of salt and pepper noise may not be fully explored. In the conventional denoising model, the $l_0$ norm and $l_1$ norm are often employed to express the sparsity of sparse variables [21]. The $l_0$ norm-based denoising methods are nondeterministic polynomial hard problems. The $l_1$ norm is the convex relaxation of the $l_0$ norm and the capacity of the $l_1$ norm to depict sparse variables is limited. What's more, the $l_1$ norm may not be adjusted by the sparsity of the depicted variables.

Second, these methods regard the contaminated image as the sum of the reconstructed image and noise and do not further refine the reconstructed image into low- and high-frequency components. Thus, it is easy to blur the high-frequency components when restoring the low-frequency components of the picture. By contrast, it is difficult to remove noise when restoring the high-frequency components of the image.

To relieve the first limitation, researchers introduced $l_p$ quasi-norm [32, 33] to further express the sparsity degree of salt and pepper noise. For instance, Wang adopted the $l_p$ quasi-norm to depict the sparsity of impulse noise and proposed a denoising method based on low-order overlapping group sparsity with $l_p$ quasi-norm and achieved promising denoising performance [19]. However, the location of the noise is still ignored.

To solve the second limitation, researchers introduced the morphological component analysis (MCA) framework [34-38] to SR for image processing. The processed image is decomposed into the cartoon, texture, and noise parts in this framework. In this way, the cartoon part containing the low-frequency components and the texture part containing the high-frequency components can be reconstructed independently using different regularization parameters. For instance, Chen *et al.* proposed an MCA-based image deblurring method [39]. Similarly, Chen *et al.* adopted the $l_0$ norm in the data fidelity term in the MCA framework to remove impulse noise in the image [40].

In the existing MCA-based denoising framework, the $l_0$ norm and $l_1$ norm are widely used for expressing the sparsity of the SR. The common sparsity regularizations in MCA-based denoising framework include discrete cosine transform [41], wavelet transform [26], curvelet transform [42], contourlet transform [43], shearlet transform [44], and so on. Although the MCA framework with SR has achieved considerable success in noise removal, there are still two issues in SR. 1) The sparsity of the sparse variables cannot be strictly controlled by a degree of freedom. 2) The dictionary or sparse transform used in SR may be unable to express the details of an image. For example, the discrete cosine transform only achieves sparse coefficients when the processed data is a periodic signal. However, the natural images in the real world rarely fit this condition. The WT involves a down-sampling operation, resulting in a missing data problem in SR.

To alleviate the first issue of SR, the $l_p$ quasi-norm is employed in MCA to improve the capacity of describing the sparse variables such as the salt and pepper noise and the coefficients of SR. In addition, since the amplitudes of salt and pepper noise are only zero or 255, it is easy to detect the amplitude of the contaminated image and find the location of the noise. After detection, the noise location information is modeled as a mask matrix, which is then employed in the proposed MCA-based denoising model to protect the clean area of the observed image and explore the location information of the noise. To alleviate the second issue of SR, the stationary Framelet transform (SFT) is adopted as the sparse transform of MCA to improve the quality of the recovered image. The SFT is a combination of stationary transform [45] and FT [46], incorporating the advantages of stationary transform and FT. Based on these motivations, an image denoising model based on SFT with $l_p$ quasi-norm (SFT_Lp) is presented.

After mathematical modeling, the alternating direction method of multipliers (ADMM) [47, 48] is adopted to solve the proposed model. Finally, we conducted experiments on several standard test images and compared them with state-of-the-art denoising methods. The peak signal-to-noise ratio (PSNR) [49], structural similarity index (SSIM) [50], and gradient magnitude similarity deviation (GMSD) [51] are used to evaluate the algorithms. The experimental results show that the proposed method outperforms the other methods. The contributions of this study are as follows:

(1) The proposed method first determines the noise location and then explores the noise location information and the sparse statistical characteristics of the salt and pepper noise to remove the noise. In this way, the noise cannot interfere with the clean area of the image during the denoising process.

(2) Compared with the current MCA framework, the SFT is adopted instead of the WT to avoid block artifacts of the WT and better represent the natural images.

(3) The $l_p$ quasi-norm is employed to better describe the sparsity of salt and pepper noise and the coefficients of the SFT. Compared with the $l_1$ norm, the $l_p$ quasi-norm has more freedom, reducing sparser solutions.

(4) The proposed model is successfully solved by using the ADMM, in which the proposed model is split into several decoupled sub-problems to solve.

The remainder of this paper is organized as follows. In section 2, the preliminaries of related work are provided for further discussion. In section 3, the proposed model and the ADMM-based solver are detailed. In section 4, we compare our experimental results with those of some state-of-the-art methods and provide the performance of the critical components in the proposed approach to verify the effectiveness of the proposed method. Finally, we discuss the main conclusions based on the experimental results.

## 2. Preliminaries

In this section, we first provide a notion list used in this paper and then review the $l_p$ quasi-norm to illustrate its strong capacity to depict the sparse variables. Then, we present the preliminaries of two-dimensional SFT to discuss the advantages of SFT. After introducing the two signal representation tools, we discuss the motivation of the proposed method.

### 2.1 Notions for symbols

In this section, we define different types of variables by using different symbols. To further help the readers to understand the proposed method, we provide the nomenclature list in Table 1.

Table 1 Nomenclature list

| Symbol | Meaning |
| --- | --- |
| $a$ | the lower-case italic letter means a scalar |
| $\boldsymbol{a}$ | the italic, bold lower-case letter means a vector form variable |
| $\boldsymbol{A}$ | the italic, bold capital letter means a matrix form variable |
| $\boldsymbol{A}(:)$ | italic, bold capital letter with $(:)$ means the vectorization form of a matrix |
| $*$ | convolution operator |
| $\circ$ | element-wise multiplication operator |
| $\boldsymbol{D}$ | stationary Framelet transform operator |
| $\boldsymbol{D}^{-1}$ | inverse stationary Framelet transform operator |
| $\|\boldsymbol{A}\|_1 = \sum_{i=1}^{N}\sum_{j=1}^{N}\|A_{ij}\|$ | $l_1$ quasi-norm of $\boldsymbol{A}$ |
| $\|\boldsymbol{A}\|_p^p = \sum_{i=1}^{N}\sum_{j=1}^{N}\|A_{ij}\|^p, p \in (0,\ 1)$ | $l_p$ quasi-norm of $\boldsymbol{A}$ |

### 2.2 Non-convex sparsity regularization depicted by $l_p$ quasi-norm

Figure 1 shows the contours of the $l_2$ norm, $l_1$ norm, and $l_p$ quasi-norm. As shown in Figure 1, the contour of the $l_p$ quasi-norm is closer to the axis of the coordinates, reducing a sparser solution than that of the $l_2$ and $l_1$ norms. Thus, the advantages of the $l_p$ quasi-norm are as follows: 1) the capacity of a sparse description is high, and 2) the $l_p$ quasi-norm has more freedoms than the $l_1$ and $l_2$ norms for one can select any $p \in (0,\ 1)$ to fit the degree of variable sparsity.

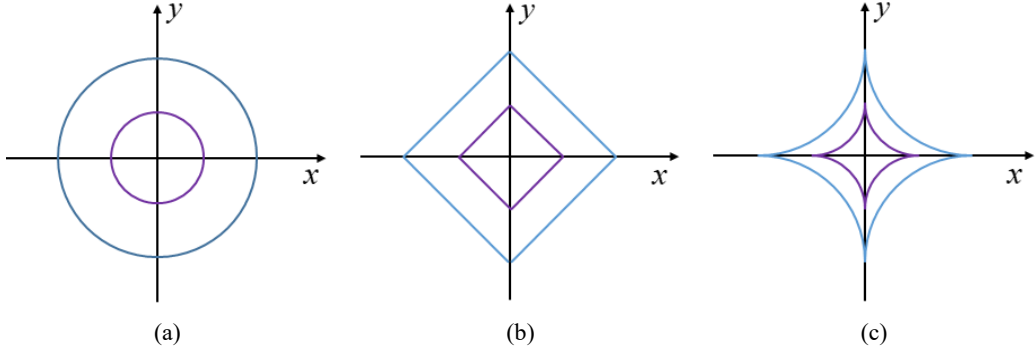

Figure 1 Contours of various norms. (a) $l_2$ norm; (b) $l_1$ norm; (c) $l_p$ quasi-norm.

Considering these advantages, we adopt the $l_p$ quasi-norm in the data fidelity term of the MCA framework to express the statistical characteristics of salt and pepper noise. Because the freedom of the $l_p$ quasi-norm lies on the parameter $p$, it seems easy to control the sparsity of the noise variable when the noise level changes. When the noise pollution level is high, the noise variable may not be highly sparse. In this case, we set the parameter $p$ to a value close to 1. Conversely, when the noise pollution level is low, the noise variable is sparser than the former. Thus, the parameter $p$ can be set to a small value close to 0. Therefore, the $l_p$ quasi-norm can fit the noise level, which helps adjust the denoising performance in the noise removal process.

**2.3 Two-dimensional stationary Framelet transform**

The SFT is the FT without a down-sampling operation. Thus, the SFT coefficients are the same as those of the original signal. These redundant coefficients ensure the time-invariant property and avoid missing information in the abrupt region of the processed signal. Thus, the recovery signal via the inverse SFT is free from the block artifacts of the traditional WT. In addition, the FT has one scaling function and two wavelet functions, whereas the WT has only one scaling function and one wavelet function. Therefore, the FT can explore more information than the WT.

Figure 2 shows a schematic sketch of the two-dimensional SFT. In Figure 2(a), the input image, $F$, is convoluted by three analytical filters: $\tilde{h} \left[\frac{1}{4}, \frac{1}{2}, \frac{1}{4}\right]^T$, $\tilde{h} \left[\frac{1}{4}, -\frac{\sqrt{2}}{4}\right]^T$, and $\tilde{h} \left[\frac{1}{4}, \frac{1}{2}, -\frac{1}{4}\right]^T$, where $\tilde{h}$ is a low-pass filter, and $\tilde{h}$, $\tilde{h}$ are high-pass filters. Then, the three convolution results are convoluted by $\tilde{h}$, $\tilde{h}$, and $\tilde{h}$. The final convolution results, $F_{LL}$, $F_{H_1L}$, $F_{H_2L}$, $F_{LH_1}$, $F_{H_1H_1}$, $F_{H_2H_1}$, $F_{LH_2}$, $F_{H_1H_2}$, and $F_{H_2H_2}$, are the results of the SFT. Figure 2(b) shows the inverse SFT. $\hat{F}$ represents the reconstructed image by inverse SFT, which is the merge-sum of SFT results convoluted with $h_0^T = \left[\frac{1}{4}, \frac{1}{2}, \frac{1}{4}\right]$, $h_1^T = \left[-\frac{\sqrt{2}}{4}, 0, \frac{\sqrt{2}}{4}\right]$, $h_2^T = \left[-\frac{1}{4}, \frac{1}{2}, -\frac{1}{4}\right]$, $h_0 = \left[\frac{1}{4}, \frac{1}{2}, \frac{1}{4}\right]^T$, $h_1 = \left[-\frac{\sqrt{2}}{4}, 0, \frac{\sqrt{2}}{4}\right]^T$, and $\boldsymbol{h}_2 = \left[ -\frac{1}{4}, \frac{1}{2}, -\frac{1}{4} \right]^T$. For example, if the processed image $\boldsymbol{F} = \begin{pmatrix} 1 & 0 & 1 & 0 & 1 \\ 0 & 0 & 0 & 0 & 0 \\ 1 & 0 & 1 & 0 & 1 \\ 0 & 0 & 0 & 0 & 0 \\ 1 & 0 & 1 & 0 & 1 \end{pmatrix} \in \mathbb{R}$, it will be convoluted by $\tilde{\boldsymbol{h}}\ \tilde{\boldsymbol{h}}$ under the reflecting boundary condition while the size of the convolution result is the same as that of $\boldsymbol{F}$. Take $\boldsymbol{F} * \tilde{\boldsymbol{h}}$ for instance, $\boldsymbol{F} * \tilde{\boldsymbol{h}} \begin{pmatrix} 0.75 & 0 & 0.75 & 0 & 0.75 \\ 0.5 & 0 & 0.5 & 0 & 0.5 \\ 0 & 0.5 & 0 & 0.5 \\ 0.5 & 0 & 0.5 & 0 & 0.5 \\ 0.75 & 0 & 0.75 & 0 & 0.75 \end{pmatrix}$. The

convolution results are then convoluted by $\tilde{\boldsymbol{h}}\ \tilde{\boldsymbol{h}}$. In this way, the SFT is calculated. The inverse SFT is calculated similarly.

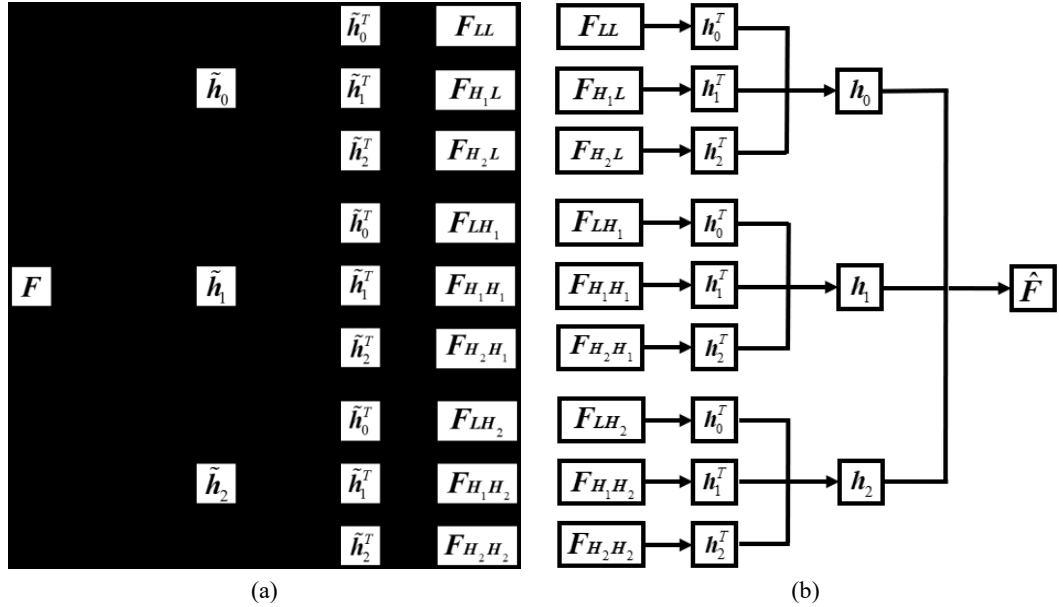

(a)    (b)

Figure 2 Two-dimensional stationary Framelet transform and inverse transform. (a) Two-dimensional stationary Framelet transform; (b) Two-dimensional inverse SFT.

As the SFT can sparsely represent natural images without the missing data problem found in WT, it is more suitable to use the SFT in MCA. Cooperation with the $l_p$ quasi-norm allows the noise sparsity and SFT coefficients to be adequately depicted, thus achieving a better-reconstructed image.

## 3. Method

In this section, we first discuss the advantages of the MCA framework and then present the improved MCA-based model developed in this study. Next, the ADMM is employed to find the solution to the proposed model. Thus, the proposed model is reduced to several simple sub-problems, which can be easily calculated.

### 3.1 Proposed denoising model via morphological component analysis with $l_p$ quasi-norm

MCA [35-38] decomposes the processed image into cartoon and texture parts by utilizing SR. The cartoon part contains the low-frequency components of the image, and the texture part contains the high-frequency components of the picture. In this way, the outline and details of the image can be recovered independently. Figure 3 illustrates the principle of MCA, where $D$ denotes the two-dimensional SFT shown in Figure 2; $N$ refers to the noise part; $F_C$ and $F_T$ represent the cartoon and texture parts, respectively. To fully explore the location information of the noise, we determine the position of the noise via the unique amplitude value of salt and pepper noise.

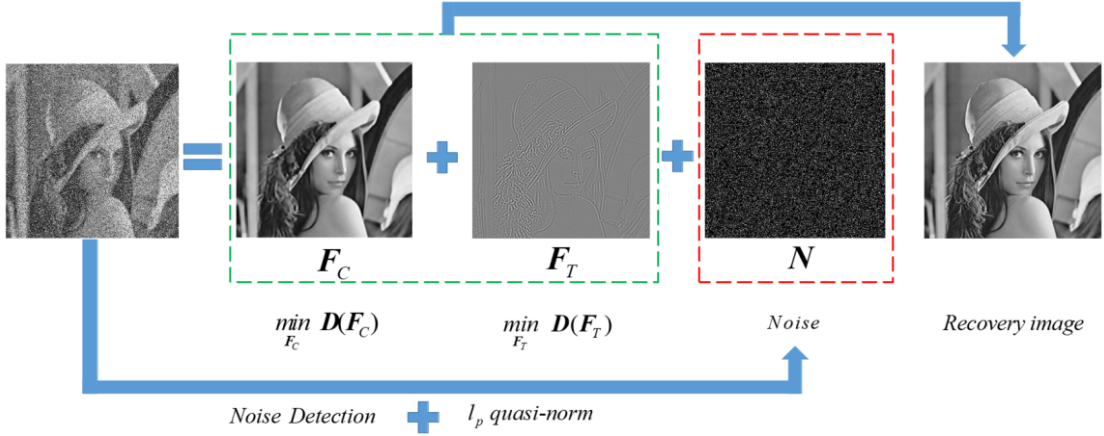

Figure 3 Diagrammatic sketch of the proposed model

Figure 3 illustrates the main concept of the proposed denoising model. In the proposed model, the processed image is considered as the sum of the cartoon, texture, and noise components. The cartoon and texture components are reconstructed separately using the SFT in the MCA framework. The unique amplitude of the salt and pepper noise makes it simple to locate the noise using the amplitude information. Thus, a noise detection operation is adopted to determine the position of the noise. In addition, the $l_p$ quasi-norm is adopted to depict the statistical characteristics of salt and pepper noise.

As discussed earlier, we adopt the MCA framework with the non-convex $l_p$ quasi-norm and SFT to improve the quality of the recovery image. The proposed model can be expressed as follows:

$$[F_T, F_C] = \underset{[F_T, F_C]}{\mathrm{argmin}}\, \alpha_0 \| M \circ F\ F - G \|_{p_0}^{p_0} + \alpha_1 \| D(F_C) \|_{p_1}^{p_1} + \alpha_2 \| D(F_T) \|_{p_2}^{p_2}, \qquad (1)$$

where $G$ is the observed image; $\alpha_0$ is the regularization parameter of the data fidelity $\| M \circ F\ F - G \|_{p_0}^{p_0}$; $\alpha_1$ and $\alpha_2$ are the regularization of the sparse prior of $F_C$ and $F_T$, respectively; and $M$ is the mask matrix with the entries $M(i,j) = \begin{cases} 1, & \text{if } 0 < G(i,j) < 255 \\ 0, & \text{if } G(i,j) = 0 \\ 0, & \text{if } G(i,j) = 255 \end{cases}$; $D$ refers to a two-dimensional SFT operator; $p_0$, $p_1$, and $p_2$ are the parameters of the $l_p$ quasi-norm. $p_0$ is used to describe the sparsity of salt and pepper noise. $p_1$ and $p_2$ are adopted to express the sparsity of the FT coefficients of $F_C$ and $F_T$, respectively.

As $F_C$ and $F_T$ are reconstructed independently, they do not interfere with one another during the recovery process. In this way, the final recovery image, $F_C + F_T$, can preserve the image detail and drastically remove the salt and pepper noise.

**3.2 Solver for the proposed model**

To solve the proposed model, the ADMM is used, where $Q_0 = M \circ F \quad F \quad -G$, $Q_1 = D(F_C)$, and $Q_2 = D(F_T)$. We introduce dual variables $\tilde{Q}$, $\tilde{Q}$, and $\tilde{Q}$ of $Q_0$, $Q_1$, and $Q_2$; the augmented Lagrange function of the proposed model is as follows:

$$J = \max_{\tilde{Q}\ \tilde{Q}} \min_{F, F_T, Q_0 - Q_2} \{\alpha_0 \|Q_0\|_{p_0}^{p_0} + \alpha_1 \|Q_1\|_{p_1}^{p_1} + \alpha_2 \|Q_1\|_{p_2}^{p_2} - \langle \lambda_0 \tilde{Q}\ Q \quad M \circ F \quad F \quad -G] \rangle$$
$$+ \frac{\lambda_0}{2} \|Q_0 - [M \circ F \quad F \quad -G]\|_2^2 - \langle \lambda_1 \tilde{Q}\ Q \quad D\ F_C) \rangle + \frac{\lambda_1}{2} \|Q_1 - D(F_C)\|_2^2 \quad (2)$$
$$- \langle \lambda_2 \tilde{Q}\ Q \quad D(F_T) \rangle + \frac{\lambda_2}{2} \|Q_2 - D(F_T)\|_2^2 \}\},$$

where $\lambda_0$, $\lambda_1$, and $\lambda_2$ represent the coefficients of quadratic penalty terms $\|Q_0 - [M \circ F \quad F \quad -G]\|_2^2$, $\|Q_1 - D(F_C)\|_2^2$, and $\|Q_2 - D(F_T)\|_2^2$, respectively.

To find the solution of (2), we solve the following subproblems for each variable separately.
(1) $F_C$ sub-problem
$F_C$ satisfies the following equation (see Appendix A):

$$\lambda_0 M \circ F \quad F \quad M \circ G \quad M \circ Q \quad Q \quad D\ Q \quad Q \quad M \circ F. \quad (3)$$

Let $X = F_C$, $A(X) = \lambda_0 M \circ X \quad X$, and $B = \lambda_0 M \circ G \quad M \circ Q \quad Q \quad D\ Q \quad Q \quad M \circ F$; then, equation (3) becomes $A(X) = B$. It is easy to solve the $F_C$ sub-problem by using the conjugate gradient method (CGM) [52]. The detail of the CGM is presented in Algorithm 1.

(2) $F_T$ sub-problem
$F_T$ satisfies the following equation (see Appendix B):

$$\lambda_0 M \circ F \quad F \quad M \circ G \quad M \circ Q \quad Q \quad D\ Q \quad Q \quad M \circ F \quad (4)$$

Similar to the $F_C$ sub-problem, equation (4) can be efficiently solved by applying Algorithm 1.

---

**Algorithm 1:** Algorithm of CGM

**Input:** operator $A$, righthand term $B$, the maximum number of iterations $K$.

**Output:** resolution $X$.

**Initialize:** initial solution $X^{(0)} = 0$, the initial residual $R^{(0)} = B \quad AX$ ).

1: Compute $P^{(0)} = R^{(0)}$;
2: For $k = 0, 1, \cdots$ do
3:      If $P^{(k)} = 0$
4:          Return $X^{(0)}$.
5:      Else
6:          $AP^{(k)} = A(P^{(k)})$
6:          $a^{(k)} = \frac{[R^{(k)}(:)]^T R^{(k)}(:)}{[P^{(k)}(:)]^T AP^{(k)}(:)}$;

---

| 7: | $X^{(k+1)} = X^{(k)} + a^{(k)} P^{(k)}$; |
|---|---|
| 8: | $R^{(k+1)} = R^{(k)} - a^{(k)} AP^{(k)}$; |
| 9: | $b^{(k)} = \dfrac{[R^{(k+1)}(:)]^T R^{(k+1)}(:)}{[R^{(k)}(:)]^T R^{(k)}(:)}$; |
| 10: | $P^{(k+1)} = R^{(k+1)} + b^{(k)} P^{(k)}$; |
| 11: | End |
| 12: | If $\dfrac{\|X^{(k+1)} - X^{(k)}\|_2}{\|X^{(k)}\|_2} < tol$ |
| 13: | break; |
| 14: | End |
| 15: End | |
| 16: Return $X^{(k+1)}$. | |

$tol=10^{-4}$ denotes the iterative threshold of the CGM algorithm.

(3) $Q_0$ sub-problem

The $Q_0$ sub-problem is as follows:

$$J_{Q_0} = \min_{Q_0} \left\{ \alpha_0 \|Q_0\|_{p_0}^{p_0} - \langle \lambda_0 \tilde{Q} \quad Q \quad M \circ F \quad F \quad G \quad \hat{} \quad Q \quad M \circ F \quad F \quad G \right\} \tag{5}$$

The method of completing the square can be applied to transform equation (5) to (the detail of the formula derivation can be found in Appendix C)

$$Q_0 = \underset{Q_0}{\operatorname{argmin}} \left\{ \alpha_0 \|Q_0\|_{p_0}^{p_0} + \dfrac{\lambda_0}{2} \|Q_0 - [M \circ F \quad F_C^{(k+1)}) - G] - \tilde{Q} \right\}_2 \tag{6}$$

Equation (6) can be solved by the generalized shrinkage discussed in reference [32], that is,

$$Q_0^{(k+1)} = \max\left\{|\Theta_0| - \left(\dfrac{\lambda_0}{\alpha_0}\right)^{p_0-2} |\Theta_0|^{p_0-1}, 0\right\} \dfrac{\Theta_0}{|\Theta_0|}, \tag{7}$$

where $\Theta_0 = M \circ F \quad F_C^{(k+1)}) - G + \tilde{Q}$.

(4) $Q_1$ sub-problem

The $Q_1$ sub-problem is as follows:

$$J_{Q_1} = \min_{Q_1} \left\{ \alpha_1 \|Q_1\|_{p_1}^{p_1} - \langle \lambda_1 \tilde{Q} \quad Q \quad D(F_C^{(k+1)}) \rangle + \dfrac{\lambda_1}{2} \|Q_1 - D(F_C^{(k+1)})\|_2^2 \right\}. \tag{8}$$

The method of completing the square, such as the derivation in Appendix C, can be used to transform equation (8) to

$$Q_1 = \underset{Q_1}{\operatorname{argmin}} \left\{ \alpha_1 \|Q_1\|_{p_1}^{p_1} + \dfrac{\lambda_1}{2} \|Q_1 - D(F_C^{(k+1)}) - \tilde{Q} \right\}_2 \tag{9}$$

According to the $l_p$ shrinkage, we have

$$Q_1^{(k+1)} = shrink_{p_1}\left(\Theta_1, \frac{\alpha_1}{\lambda_1}\right) = max\left\{|\Theta_1| - \left(\frac{\lambda_1}{\alpha_1}\right)^{p_1-2}|\Theta_1|^{p_1-1}, 0\right\}\frac{\Theta_1}{|\Theta_1|}, \quad (10)$$

where $\Theta_1 = D(F_C^{(k+1)}) + \tilde{Q}$.

(5) $Q_2$ sub-problem

The $Q_2$ sub-problem is as follows:

$$J_{Q_2} = \alpha_2 \|Q_2\|_{p_2}^{p_2} - \langle \lambda_2 \tilde{Q}, Q_2 - D(F_T^{(k+1)})\rangle + \frac{\lambda_2}{2}\|Q_2 - D(F_T^{(k+1)})\|_2^2. \quad (11)$$

By using the method of completing the square, such as the derivation in Appendix C, we have

$$Q_2 = \underset{Q_2}{argmin}\left\{\alpha_2\|Q_2\|_{p_2}^{p_2} + \frac{\lambda_2}{2}\|Q_2 - D(F_T^{(k+1)}) - \tilde{Q}\|_2^2\right\} \quad (12)$$

According to $l_p$ shrinkage, we have

$$Q_2^{(k+1)} = shrink_{p_2}\left(\Theta_2, \frac{\alpha_2}{\lambda_2}\right) = max\left\{|\Theta_2| - \left(\frac{\lambda_2}{\alpha_2}\right)^{p_2-2}|\Theta_2|^{p_2-1}, 0\right\}\frac{\Theta_2}{|\Theta_2|}, \quad (13)$$

where $\Theta_2 = D(F_T^{(k+1)}) + \tilde{Q}$.

(6) $\tilde{Q}$ sub-problem

The $\tilde{Q}$ sub-problem is as follows:

$$J_{\tilde{Q}} = \underset{\tilde{Q}}{argmin}\left\{-\langle \lambda_0 \tilde{Q}, Q_0 - [M \circ F - F_C^{(k+1)}) - G]\rangle\right\}. \quad (14)$$

By using the gradient ascend method, we have

$$\tilde{Q} = \tilde{Q} + \gamma[M \circ F - F_C^{(k+1)}) - G - Q_0^{(k+1)}], \quad (15)$$

where $\gamma$ represents the learning ratio.

(7) $\tilde{Q}$ sub-problem

The $\tilde{Q}$ sub-problem is as follows:

$$J_{\tilde{Q}} = \underset{\tilde{Q}}{argmin}\left\{-\langle \lambda_1 \tilde{Q}, Q_1 - D(F_C^{(k+1)})\rangle\right\}. \quad (16)$$

By using the gradient ascend method, we have

$$\tilde{Q} = \tilde{Q} + \gamma[D(F_C^{(k+1)}) - Q_1^{(k+1)}]. \quad (17)$$

(8) $\tilde{Q}$ sub-problem

The $\tilde{Q}$ sub-problem is as follows:

$$J_{\tilde{Q}} = \underset{\tilde{Q}}{argmin}\left\{-\langle \lambda_2 \tilde{Q}, Q_2 - D(F_T^{(k+1)})\rangle\right\}. \quad (18)$$

By using the gradient ascend method, we have

$$\tilde{Q} = \tilde{Q} + \gamma[D(F_T^{(k+1)}) - Q_2^{(k+1)}]. \quad (19)$$

The proposed method is summarized as Algorithm 2.

---
**Algorithm 2**: Proposed method
---
**Input:** observed image $G$.

**Output:** recovered image $F = F_C + F_T$.

**Initialize:** $F_C^{(0)}$, $F_T^{(0)}$ $F^{(0)}$, $Q_0^{(0)}$, $Q_1^{(0)}$, $Q_2^{(0)}$, $\tilde{Q}$, $\tilde{Q}$, $\tilde{Q}$, maximum number of iterations $K$.

**1: For** $k = 0, 1, \cdots$ **do**;

**2:**   Compute $F_C^{(k+1)}$ by solving equation (3) using **Algorithm 1**;

**3:**   Compute $F_T^{(k+1)}$ by solving equation (4) using **Algorithm 1**;

**4:**   $\hat{F}^{(k+1)} = F_C^{(k+1)} + F_T^{(k+1)}$;

**5:**   Update $Q_0^{(k+1)}$ by equation (7);

**6:**   Update $Q_1^{(k+1)}$ by equation (10);

**7:**   Update $Q_2^{(k+1)}$ by equation (12);

**8:**   Update $\tilde{Q}$ by equation (15);

**9:**   Update $\tilde{Q}$ by equation (17);

**10:**  Update $\tilde{Q}$ by equation (19);

**11:**  If $\dfrac{\left\| F^{(k+1)} - F^{(k)} \right\|_2}{\left\| F^{(k)} \right\|_2} < tol$;

**12:**      break;

**13:**  End

**14: End**

**15: Return** $\hat{F}^{(k+1)}$.
---

where the maximal iteration number, $K$, is set as 100, and the iteration stop threshold, $tol$, is set as $10^{-4}$.

## 4. Experiments

This section presents several experiments to verify the proposed method. We introduce the primary measure indexes used in this study, compare the proposed method with some state-of-the-art methods, and demonstrate the effectiveness of the critical components of the proposed method by using ablation experiments.

### 4.1 Algorithm evaluation indexes and parameter setting

We adopted some common image evaluation indexes to estimate the quality of the recovered image using different algorithms. The evaluation indexes used in the experiments were the PSNR [49], SSIM [50], and GMSD [51].

The PSNR [49] is the most common and widely used objective measurement of image recovery, and it is computed as follows:

$$PSNR(X,Y) = 10\lg \frac{(max(X))^2}{\frac{1}{N^2}\sum_{i=1}^{N}\sum_{j=1}^{N}(X_{ij}-Y_{ij})^2}, \quad (20)$$

where $X \in \mathbb{R}$ represents the original image and $Y \in \mathbb{R}$ represents the recovered image. A high PSNR value indicates that the reconstructed image is similar to the original image.

The SSIM is an objective measurement of image recovery, and it is defined as

$$SSIM(X,Y) = \frac{(2u_X u_Y + (Lk_1)^2)(2\sigma_{XY} + (Lk_2)^2)}{(u_X^2 + u_Y^2 + (Lk_1)^2)(\sigma_X^2 + \sigma_Y^2 + (Lk_2)^2)}, \quad (21)$$

where $u_X$ is the mean value of $X$, $u_Y$ is the mean value of $Y$, $\sigma_X^2$ is the variance of $X$, $\sigma_Y^2$ is the variance of $Y$, $\sigma_{XY}$ is the covariance of $X$ and $Y$, $L$ is the maximum gray value of $X$, and $k_1$, $k_2$ are the parameters maintaining the denominator as a non-zero number. The SSIM should be within the range $[0,1]$. When the reconstructed image is similar to the original image, the SSIM will be close to 1.

The GMSD is another objective measurement of image recovery, and it is defined as

$$GMSD = \sqrt{\frac{1}{MN}\sum_{i=1}^{M}\sum_{j=1}^{N}\left\{[GMS]_{i,j} - \frac{1}{MN}\sum_{i=1}^{M}\sum_{j=1}^{N}[GMS]_{i,j}\right\}^2}, \quad (22)$$

where $[GMS]_{i,j}$ represents the local gradient magnitude similarity concerning the pixels in the $i$-th row and $j$-th column in the processed image. It is defined as follows:

$$[GMS]_{i,j} = \frac{2[m_X]_{i,j}[m_Y]_{i,j} + c}{[m_X]_{i,j}^2 + [m_Y]_{i,j}^2 + c}, \quad (23)$$

where $c$ is a constant that keeps the denominator as a nonzero number. $[m_X]_{i,j}$ and $[m_Y]_{i,j}$ represent the gradient magnitudes of $X$ and $Y$ at the $i$-th row and $j$-th column in the processed image, respectively, and these are defined as follows:

$$[m_X]_{i,j} = \sqrt{[X*H_1]_{i,j}^2 + [X*H_2]_{i,j}^2}, \quad [m_Y]_{i,j} = \sqrt{[Y*H_1]_{i,j}^2 + [Y*H_2]_{i,j}^2}, \quad (24)$$

where $H_1 = \begin{bmatrix} 1/3 & 0 & 1/3 \\ 1/3 & 0 & 1/3 \\ 1/3 & 0 & 1/3 \end{bmatrix}$ and $H_2 = \begin{bmatrix} 1/3 & 1/3 & 1/3 \\ 0 & 0 & 0 \\ 1/3 & 1/3 & 1/3 \end{bmatrix}$.

When the value of the GMSD is small, the deviation of the gradient magnitude similarity between the reconstructed image and the original image is small. Thus, the smaller the GMSD, the higher is the quality of the recovered image.

We selected eight standard images, shown in Figure 4, to verify the effectiveness of the proposed model. To compare the proposed model with other methods efficiently, the images were down-sampled to 256×256; only Lena512 (Figure 4(f)) had the size of 512×512.

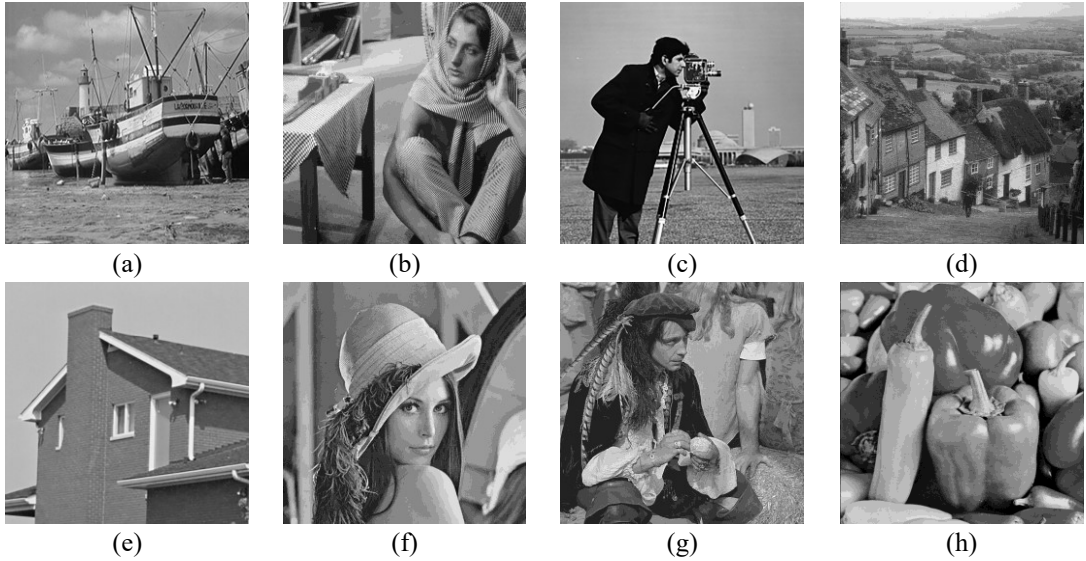

Figure 4 Standard test images. (a) Boat; (b) Barbara; (c) Camera; (d) Gold hill; (e) House; (f) Lena512; (g) Man; (h) Peppers.

In the subsequent experiment, the parameter values are selected as follows: $p_0$, $p_1$, $p_2 \in (0, 1)$. These parameters should be adjusted to achieve the best denoising performance. The parameter $p_0$ is set to depict the degree of noise sparsity. Typically, when the noise level is low, $p_0$ should be adjusted to be near 0. By contrast, when the noise level is high, i.e., the noise is not sparse enough, $p_0$ should be close to 1; Parameter $p_1$ is set to describe the sparsity of the cartoon part in the Framelet domain and should be set larger than $p_2$, which controls the sparsity of texture part in the Framelet domain because we assume that the texture parts are more sparse than the cartoon parts; $\alpha_0 = 1$; $\alpha_1$, $\alpha_2$ were set in the range $[0, 10]$. The parameter $\alpha_1$ is used to control the sparsity of cartoon parts in the Framelet domain and $\alpha_2$ is set for the sparsity of texture parts in the Framelet domain. We assume that the texture parts are sparser than the cartoon parts. Thus, we arranged $\alpha_1 = \frac{1}{2}\alpha_2$ for convenience; $\lambda_0$, $\lambda_1$, $\lambda_2$ are set in the range $[0, 1]$. The parameter $\gamma$ is set as $\gamma = 10^{-6}$.

**4.2 Comparison with some TV-based methods and convergence verification**
**4.2.1 Comparison with some TV-based methods**

In this section, we compare the proposed method (SFT_Lp) with some TV-based methods, which are based on anisotropic total variation (ATV) [18], isotropic total variation (ITV) [53], ITV with $l_p$ quasi-norm (ITV_Lp), low-order overlapping group sparsity with $l_p$ quasi-norm (LOGS_Lp) [19], and high-order overlapping group sparsity [54] with $l_p$ quasi-norm (HOGS_Lp).

To determine the reconstructed image quality, we used the proposed model and the ATV, ITV, ITV_Lp, LOGS_Lp, and HOGS_Lp models to restore the "Lena512" image for local magnification. The results are shown in Figure 5. As seen in the figure, an eye and an eyebrow of Lena are perfectly reconstructed by the proposed model, whereas the compared methods fail to recover these details. The ITV_Lp denoising model outperforms the ITV denoising model because it benefits from using $l_p$ quasi-norm instead of the $l_1$ norm in the data fidelity term; the $l_p$ quasi-norm can depict the statistical characteristics of the sparse variable more precisely than the $l_1$ norm. The ITV_Lp, LOGS_Lp, HOGS_Lp, and proposed methods employ the $l_p$ quasi-norm in the data fidelity. Under these conditions, we found that the stationary Framelet regularization in the proposed model outperforms that in the ITV_Lp, LOGS_Lp, and HOGS_Lp models.

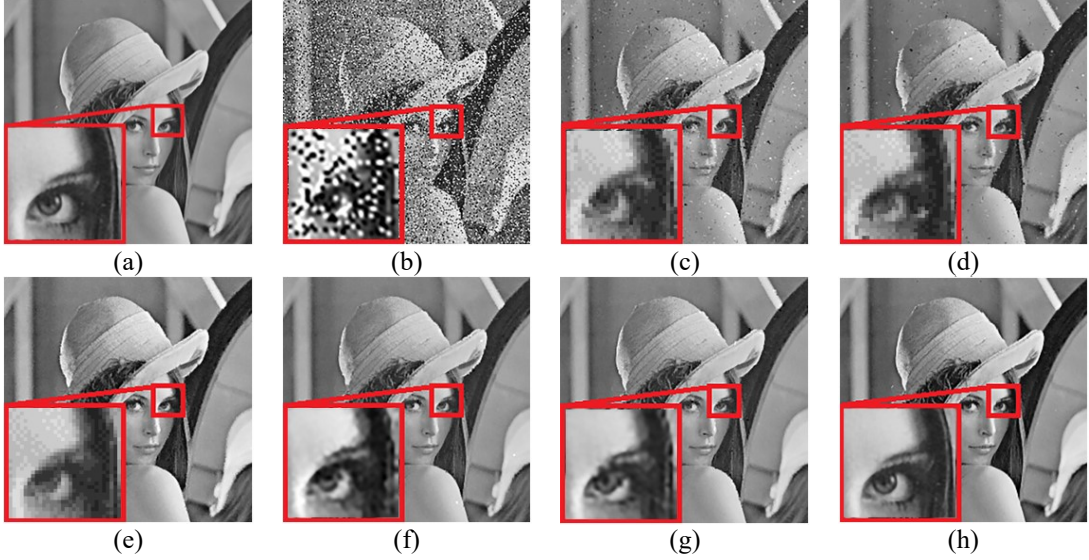

Figure 5 Partial enlarged view of the "Lena512" under 20% salt and pepper noise. (a) Original image of "Lena512"; (b) "Lena512" under 20% noise level interference; (c) ATV method (PSNR=31.74 dB, SSIM=0.954, and GMSD=0.056); (d) ITV method (PSNR=31.82 dB, SSIM=0.956, and GMSD=0.054); (e) ITV_Lp method (PSRN=35.63 dB, SSIM=0.987, and GMSD=0.020); (f) LOGS_Lp method (PSNR=38.82 dB, SSIM=0.994, and GMSD=0.013); (g) HOGS_Lp method (PSNR=39.33 dB, SSIM=0.994, and GMSD=0.011); (h) proposed method (PSNR=**42.18 dB**, SSIM=**0.996**, and GMSD=**0.005**).

### 4.2.2 Convergence verification

Although the sparsity regularization depicted by $l_p$ quasi-norm is a non-convex sparsity regularization, the guarantee of the iterative p-shrinkage convergence has been proved in [33]. In this section, we provide a convergence verification by three dynamic iteration curves shown in Figure 6.

For a fair comparison, the stop iteration threshold parameter, $tol$, of all the compared methods is set as $10^{-4}$, and the parameters of the compared methods are all adjusted by hand to achieve their best performances.

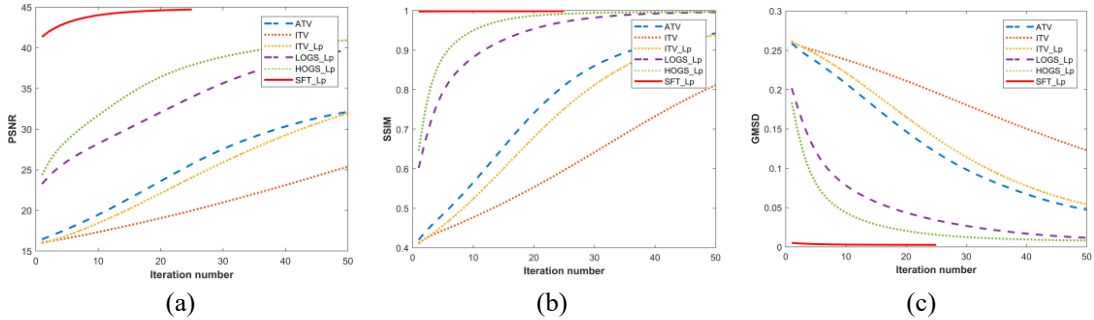

Figure 6 Dynamic iterative diagrams of the PSNR, SSIM, and GMSD values of Lena512 by different methods. Dynamic iterative graph of the (a) PSNR value under 10% salt and pepper noise; (b) SSIM value under 10% salt and pepper noise; (c) GMSD value under 10% salt and pepper noise.

As is seen in Figure 6, The whole iteration number of the proposed method is only 25, while the other compared methods do not converge before 50. This indicates the proposed model (SFT_Lp) quickly converges to the final solution and outperforms the compared algorithms. Besides, the dynamic, iterative curves show that the proposed method can converge to better indexes stably compared with the other methods.

**4.3 Comparison with some median filter-based methods**

In this section, we compare some median-filter-based methods, namely, the adaptive frequency median filter (AFMF) [55] method, iterative mean filter (IMF) [56] method, different applied median filter (DAMF) [57] method, two-stage filter (TSF) [58] method, and adaptive Riesz median filter (ARMF) [59] method with our proposed model. Table 2–5 list the experiment results. The Lena image with a size of 512×512 is denoted as "Lena512," and the remaining images in the tables were down-sampled to 256×256. The best performances are in bold font. The denoising performance indicated in Tables 2–4 shows that the proposed model outperforms the median-filter-based methods in most situations in terms of the PSNR, SSIM, and GMSD. However, the cost time shown in Table 5 indicates that the proposed method is time-consuming compared with the other methods.

To further illustrate the advantage of the proposed model, we show the enlarged, reconstructed image in Figure 7. Figure 7 (h) depicts that the peppers' edge can be recovered by the proposed method. However, the edges recovered by the median-filter-based methods are still polluted by the noise. Thus, we can conclude that the proposed method outperforms the compared methods in quantitative and qualitative comparisons.

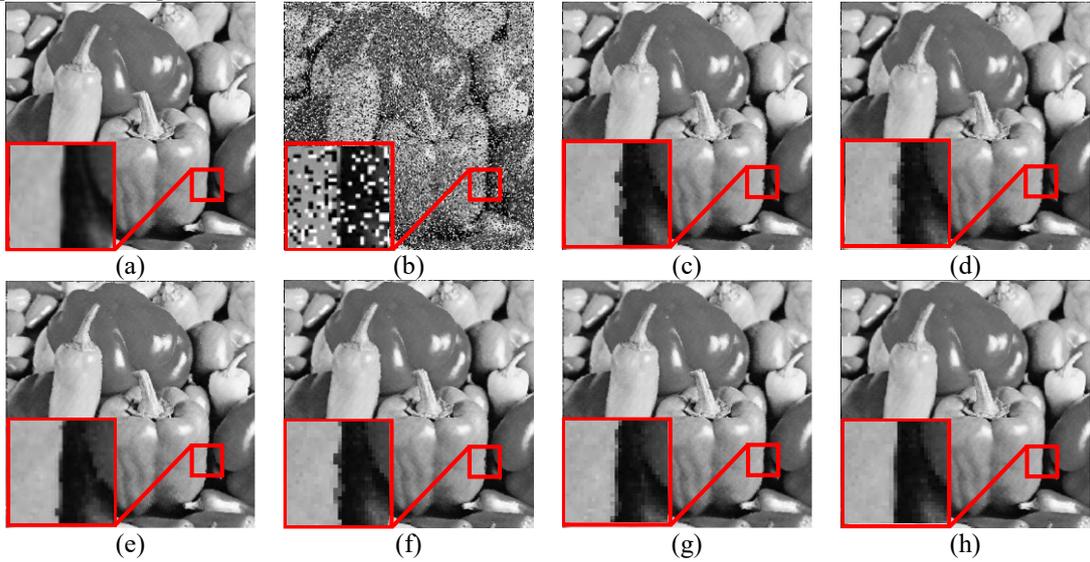

Figure 7 A partially enlarged view of the "Peppers" under 30% salt and pepper noise. (a) Original image of "Peppers"; (b) "Peppers" under 30% noise level interference; (c) AFMF method; (d) IMF method; (e) DAMF method; (f) TSF method; (g) ARMF method; (h) proposed method.

Table 2 PSNR (dB) of different methods under different level noise

| Image | Method | 10% | 20% | 30% | 40% | 50% | 60% | 70% | 80% | 90% |
|---|---|---|---|---|---|---|---|---|---|---|
| Lena512 | AFMF | 38.22 | 36.84 | 35.21 | 33.68 | 32.08 | 30.77 | 29.12 | 27.47 | 21.95 |
| | IMF | 42.33 | 39.15 | 36.92 | 35.40 | 33.99 | 32.49 | 31.16 | 29.51 | 27.44 |
| | DAMF | 43.06 | 39.17 | 36.82 | 34.75 | 33.32 | 31.62 | 30.28 | 28.47 | 25.79 |
| | TSF | 43.15 | 39.05 | 36.65 | 34.78 | 33.14 | 31.60 | 30.24 | 28.57 | 26.12 |
| | ARMF | 43.03 | 39.77 | 37.54 | 35.72 | 34.15 | 32.59 | 30.76 | 28.88 | 26.13 |
| | SFT_Lp | **45.77** | **42.18** | **39.12** | **37.26** | **35.50** | **33.72** | **32.13** | **29.94** | **27.69** |
| House | AFMF | 36.65 | 35.19 | 33.80 | 32.36 | 30.73 | 29.54 | 27.92 | 26.16 | 21.28 |
| | IMF | 40.76 | 38.01 | 35.76 | 34.24 | 32.92 | 31.55 | 30.09 | 28.36 | 26.54 |
| | DAMF | 42.36 | 38.40 | 35.40 | 33.33 | 31.82 | 30.18 | 28.99 | 27.29 | 24.69 |

| Image | Method | | | | | | | | |
|---|---|---|---|---|---|---|---|---|---|
| | TSF | 42.57 | 37.89 | 35.18 | 33.51 | 31.71 | 30.54 | 29.07 | 27.52 | 25.09 |
| | ARMF | 42.18 | 38.51 | 36.54 | 34.80 | 33.07 | 31.35 | 29.56 | 27.67 | 25.08 |
| | SFT_Lp | **46.83** | **42.91** | **39.93** | **37.95** | **36.12** | **34.61** | **32.29** | **30.08** | **27.65** |
| Gold hill | AFMF | 31.06 | 30.51 | 29.44 | 28.24 | 27.13 | 26.13 | 25.00 | 23.58 | 20.47 |
| | IMF | 36.53 | 33.16 | 31.09 | 29.79 | 28.61 | 27.51 | 26.60 | 25.35 | 23.84 |
| | DAMF | 36.65 | 32.90 | 30.81 | 29.15 | 27.86 | 26.48 | 25.48 | 24.13 | 22.59 |
| | TSF | 36.40 | 32.94 | 30.94 | 29.10 | 27.81 | 26.65 | 25.40 | 24.37 | 22.62 |
| | ARMF | 37.11 | 33.68 | 31.63 | 30.07 | 28.50 | 27.28 | 25.86 | 24.46 | 22.71 |
| | SFT_Lp | **37.69** | **34.01** | **32.16** | **31.14** | **29.74** | **28.41** | **27.05** | **25.62** | **23.92** |
| Man | AFMF | 30.32 | 29.33 | 28.66 | 27.56 | 26.33 | 25.34 | 24.10 | 22.62 | 19.86 |
| | IMF | 35.35 | 32.18 | 30.40 | 28.99 | 27.69 | 26.65 | 25.55 | 24.44 | 23.01 |
| | DAMF | 36.07 | 32.56 | 30.36 | 28.53 | 27.04 | 25.87 | 24.67 | 23.26 | 21.44 |
| | TSF | 35.88 | 32.53 | 30.35 | 28.59 | 27.25 | 25.83 | 24.61 | 23.27 | 21.80 |
| | ARMF | 36.05 | 32.61 | 30.75 | 28.87 | 27.68 | 26.43 | 25.08 | 23.55 | 21.65 |
| | SFT_Lp | **39.76** | **33.82** | **31.41** | **29.99** | **28.59** | **27.52** | **26.15** | **24.63** | **23.06** |
| Peppers | AFMF | 31.79 | 30.28 | 29.27 | 28.29 | 26.95 | 25.34 | 24.46 | 22.84 | 19.33 |
| | IMF | 36.96 | 33.86 | 31.57 | 30.02 | 28.70 | 27.48 | 26.18 | 24.81 | 23.03 |
| | DAMF | 39.05 | 34.62 | 31.36 | 29.83 | 27.91 | 26.70 | 25.29 | 23.77 | 21.65 |
| | TSF | 37.26 | 34.48 | 31.99 | 29.60 | 28.14 | 26.49 | 25.18 | 23.61 | 21.45 |
| | ARMF | 37.65 | 34.35 | 32.16 | 30.29 | 28.59 | 27.09 | 25.26 | 24.06 | 21.67 |
| | SFT_Lp | **40.66** | **37.30** | **33.44** | **32.13** | **30.51** | **28.91** | **27.07** | **24.97** | **23.15** |
| Camera | AFMF | 29.98 | 28.82 | 27.67 | 26.53 | 25.31 | 24.08 | 22.83 | 21.53 | 18.70 |
| | IMF | 35.06 | 31.76 | 29.76 | 28.16 | 26.79 | 25.79 | 24.66 | 23.46 | 21.59 |
| | DAMF | 35.06 | 31.76 | 29.71 | 27.71 | 26.34 | 24.84 | 23.70 | 22.16 | 20.27 |
| | TSF | 35.12 | 31.86 | 29.49 | 27.57 | 26.12 | 25.12 | 23.96 | 22.49 | 20.78 |
| | ARMF | 35.22 | 32.39 | 30.37 | 28.89 | 27.18 | 25.91 | 24.28 | 22.51 | 20.45 |
| | SFT_Lp | **38.33** | **34.62** | **31.41** | **29.99** | **28.51** | **26.70** | **25.26** | **23.84** | **21.68** |
| Barbara | AFMF | 26.46 | 25.87 | 25.09 | 24.11 | 23.30 | 22.35 | 21.50 | 20.65 | 18.48 |
| | IMF | 32.07 | 28.98 | 27.41 | 25.93 | 24.87 | 23.97 | 23.19 | 22.22 | 20.98 |
| | DAMF | 31.49 | 28.32 | 26.56 | 24.92 | 23.76 | 22.72 | 21.74 | 20.74 | 19.43 |
| | TSF | 31.66 | 28.56 | 26.19 | 24.82 | 23.75 | 22.67 | 21.74 | 20.93 | 19.57 |
| | ARMF | 31.98 | 28.76 | 26.79 | 25.41 | 24.22 | 23.05 | 21.93 | 20.88 | 19.54 |
| | SFT_Lp | **34.71** | **31.05** | **27.79** | **27.11** | **25.82** | **24.64** | **23.63** | **22.41** | **21.03** |
| Boat | AFMF | 29.22 | 28.48 | 27.24 | 26.41 | 25.33 | 24.08 | 22.87 | 21.79 | 19.07 |
| | IMF | 34.01 | 31.00 | 29.05 | 27.88 | 26.56 | 25.49 | 24.51 | 23.33 | 21.84 |
| | DAMF | 34.72 | 30.97 | 28.88 | 27.34 | 25.87 | 24.66 | 23.24 | 22.17 | 20.53 |
| | TSF | 34.49 | 30.96 | 28.79 | 27.36 | 25.97 | 24.69 | 23.55 | 22.35 | 20.92 |
| | ARMF | 34.60 | 31.51 | 29.67 | 28.03 | 26.58 | 25.22 | 23.96 | 22.60 | 20.70 |
| | SFT_Lp | **36.65** | **33.67** | **31.35** | **29.31** | **27.91** | **26.53** | **25.21** | **23.87** | **22.15** |

Table 3 SSIM of different methods under different level noise

| Image | Method | 10% | 20% | 30% | 40% | 50% | 60% | 70% | 80% | 90% |
|---|---|---|---|---|---|---|---|---|---|---|
| Lena512 | AFMF | 0.991 | 0.990 | 0.986 | 0.982 | 0.974 | 0.964 | 0.947 | 0.916 | 0.796 |
| | IMF | 0.997 | 0.995 | 0.991 | 0.986 | 0.979 | 0.970 | 0.955 | 0.931 | **0.884** |
| | TSF | **0.998** | 0.995 | 0.990 | 0.985 | 0.978 | 0.969 | 0.956 | 0.931 | 0.879 |
| | DAMF | 0.997 | 0.995 | 0.990 | 0.986 | 0.979 | 0.968 | 0.954 | 0.930 | 0.870 |
| | ARMF | **0.998** | 0.995 | 0.992 | 0.987 | 0.981 | 0.971 | 0.957 | 0.932 | 0.873 |
| | SFT_Lp | **0.998** | **0.996** | **0.993** | **0.988** | **0.982** | **0.973** | **0.959** | **0.933** | **0.884** |
| House | AFMF | 0.946 | 0.942 | 0.934 | 0.918 | 0.897 | 0.878 | 0.836 | 0.791 | 0.670 |
| | IMF | 0.986 | 0.973 | 0.957 | 0.941 | 0.924 | 0.901 | 0.875 | 0.834 | 0.787 |
| | DAMF | 0.987 | 0.972 | 0.954 | 0.933 | 0.909 | 0.884 | 0.849 | 0.807 | 0.743 |
| | TSF | 0.988 | 0.970 | 0.954 | 0.933 | 0.909 | 0.884 | 0.853 | 0.818 | 0.754 |
| | ARMF | 0.988 | 0.976 | 0.963 | 0.948 | 0.928 | 0.899 | 0.865 | 0.820 | 0.750 |
| | SFT_Lp | **0.993** | **0.984** | **0.969** | **0.962** | **0.947** | **0.928** | **0.903** | **0.865** | **0.795** |
| Gold hill | AFMF | 0.909 | 0.902 | 0.883 | 0.854 | 0.816 | 0.770 | 0.704 | 0.621 | 0.477 |
| | IMF | 0.977 | 0.951 | 0.922 | 0.890 | 0.855 | 0.815 | 0.762 | 0.690 | 0.591 |
| | DAMF | 0.977 | 0.948 | 0.917 | 0.879 | 0.839 | 0.787 | 0.730 | 0.654 | 0.543 |
| | TSF | 0.977 | 0.950 | 0.921 | 0.880 | 0.840 | 0.791 | 0.730 | 0.664 | 0.558 |
| | ARMF | **0.979** | 0.955 | 0.928 | 0.897 | 0.856 | 0.810 | 0.750 | 0.670 | 0.552 |
| | SFT_Lp | **0.979** | **0.956** | **0.929** | **0.912** | **0.882** | **0.839** | **0.787** | **0.701** | **0.603** |
| Man | AFMF | 0.913 | 0.906 | 0.891 | 0.869 | 0.830 | 0.786 | 0.726 | 0.644 | 0.499 |
| | IMF | 0.979 | 0.953 | 0.929 | 0.900 | 0.862 | 0.824 | 0.771 | 0.707 | **0.612** |
| | DAMF | 0.978 | 0.954 | 0.928 | 0.892 | 0.853 | 0.807 | 0.752 | 0.682 | 0.567 |
| | TSF | 0.980 | 0.956 | 0.918 | 0.893 | 0.858 | 0.810 | 0.751 | 0.679 | 0.573 |
| | ARMF | 0.981 | 0.957 | 0.933 | 0.898 | 0.864 | 0.822 | 0.771 | 0.692 | 0.576 |
| | SFT_Lp | **0.982** | **0.964** | **0.940** | **0.914** | **0.879** | **0.843** | **0.789** | **0.718** | 0.599 |
| Peppers | AFMF | 0.951 | 0.944 | 0.932 | 0.917 | 0.891 | 0.856 | 0.820 | 0.755 | 0.616 |
| | IMF | 0.986 | 0.971 | 0.956 | 0.937 | 0.915 | 0.890 | 0.858 | 0.814 | 0.741 |
| | DAMF | 0.988 | 0.972 | 0.954 | 0.931 | 0.905 | 0.875 | 0.834 | 0.783 | 0.690 |
| | TSF | 0.988 | 0.972 | 0.955 | 0.932 | 0.905 | 0.874 | 0.837 | 0.782 | 0.691 |
| | ARMF | 0.987 | 0.974 | 0.958 | 0.937 | 0.915 | 0.884 | 0.844 | 0.792 | 0.693 |
| | SFT_Lp | **0.989** | **0.978** | **0.961** | **0.948** | **0.929** | **0.906** | **0.875** | **0.828** | **0.745** |
| Camera | AFMF | 0.941 | 0.934 | 0.926 | 0.908 | 0.883 | 0.851 | 0.805 | 0.752 | 0.623 |
| | IMF | 0.986 | 0.968 | 0.951 | 0.930 | 0.903 | 0.877 | 0.841 | 0.793 | 0.720 |
| | DAMF | 0.986 | 0.969 | 0.951 | 0.926 | 0.901 | 0.866 | 0.830 | 0.778 | 0.698 |
| | TSF | 0.986 | 0.969 | 0.949 | 0.925 | 0.896 | 0.869 | 0.833 | 0.782 | 0.713 |
| | ARMF | 0.987 | 0.972 | **0.957** | 0.936 | 0.912 | 0.883 | 0.841 | 0.787 | 0.707 |
| | SFT_Lp | **0.990** | **0.979** | 0.956 | **0.945** | **0.925** | **0.892** | **0.857** | **0.809** | **0.724** |
| Barbara | AFMF | 0.898 | 0.888 | 0.867 | 0.836 | 0.801 | 0.752 | 0.696 | 0.626 | 0.490 |
| | IMF | 0.974 | 0.945 | 0.917 | 0.883 | 0.845 | 0.804 | 0.753 | 0.689 | **0.597** |
| | DAMF | 0.972 | 0.940 | 0.906 | 0.864 | 0.819 | 0.768 | 0.707 | 0.639 | 0.542 |
| | TSF | 0.972 | 0.942 | 0.902 | 0.862 | 0.820 | 0.767 | 0.709 | 0.644 | 0.543 |
| | ARMF | 0.973 | 0.943 | 0.909 | 0.871 | 0.828 | 0.774 | 0.715 | 0.642 | 0.547 |
| | SFT_Lp | **0.985** | **0.965** | **0.927** | **0.907** | **0.874** | **0.829** | **0.767** | **0.703** | 0.589 |
| Boat | AFMF | 0.900 | 0.895 | 0.877 | 0.852 | 0.812 | 0.759 | 0.692 | 0.619 | 0.469 |
| | IMF | 0.975 | 0.948 | 0.918 | 0.886 | 0.847 | 0.800 | 0.748 | 0.678 | 0.578 |
| | DAMF | 0.975 | 0.947 | 0.915 | 0.875 | 0.832 | 0.783 | 0.715 | 0.644 | 0.531 |
| | TSF | 0.975 | 0.948 | 0.917 | 0.875 | 0.834 | 0.778 | 0.723 | 0.650 | 0.544 |
| | ARMF | 0.977 | 0.952 | 0.925 | 0.894 | 0.851 | 0.802 | 0.741 | 0.662 | 0.543 |

|  | SFT_Lp | **0.982** | **0.963** | **0.928** | **0.911** | **0.880** | **0.833** | **0.772** | **0.691** | **0.582** |

Table 4 GMSD of different methods under different level noise

| Image | Method | 10% | 20% | 30% | 40% | 50% | 60% | 70% | 80% | 90% |
|---|---|---|---|---|---|---|---|---|---|---|
| Lena512 | AFMF | 0.008 | 0.011 | 0.016 | 0.022 | 0.031 | 0.040 | 0.054 | 0.080 | 0.158 |
|  | IMF | 0.003 | 0.008 | 0.012 | 0.019 | 0.027 | 0.038 | 0.051 | 0.071 | 0.102 |
|  | DAMF | 0.003 | 0.008 | 0.012 | 0.019 | 0.025 | 0.036 | 0.048 | 0.064 | 0.100 |
|  | TSF | 0.003 | 0.008 | 0.013 | 0.019 | 0.026 | 0.035 | 0.048 | 0.067 | 0.100 |
|  | ARMF | 0.003 | 0.007 | 0.011 | 0.016 | 0.024 | 0.033 | 0.045 | **0.061** | **0.095** |
|  | SFT_Lp | **0.002** | **0.005** | **0.009** | **0.015** | **0.022** | **0.032** | **0.043** | **0.061** | 0.123 |
| House | AFMF | 0.012 | 0.016 | 0.021 | 0.027 | 0.040 | 0.050 | 0.062 | 0.089 | 0.169 |
|  | IMF | 0.004 | 0.010 | 0.015 | 0.023 | 0.032 | 0.042 | 0.057 | 0.083 | **0.115** |
|  | DAMF | 0.004 | 0.009 | 0.016 | 0.023 | 0.032 | 0.045 | 0.055 | 0.076 | **0.115** |
|  | TSF | 0.003 | 0.010 | 0.017 | 0.022 | 0.034 | 0.045 | 0.055 | 0.074 | 0.120 |
|  | ARMF | 0.003 | 0.008 | 0.012 | 0.020 | 0.026 | 0.039 | 0.054 | 0.072 | 0.120 |
|  | SFT_Lp | **0.002** | **0.005** | **0.010** | **0.014** | **0.020** | **0.030** | **0.045** | **0.069** | 0.132 |
| Gold hill | AFMF | 0.026 | 0.032 | 0.038 | 0.048 | 0.056 | 0.074 | 0.091 | 0.114 | 0.175 |
|  | IMF | 0.010 | 0.022 | 0.029 | 0.039 | 0.052 | 0.063 | 0.080 | 0.104 | **0.132** |
|  | DAMF | 0.010 | 0.021 | 0.035 | 0.045 | 0.055 | 0.067 | 0.084 | 0.108 | 0.145 |
|  | TSF | 0.012 | 0.020 | 0.031 | 0.044 | 0.054 | 0.067 | 0.084 | 0.105 | 0.139 |
|  | ARMF | 0.009 | **0.017** | 0.029 | 0.035 | 0.050 | 0.059 | 0.080 | 0.104 | 0.144 |
|  | SFT_Lp | **0.008** | 0.019 | **0.024** | **0.033** | **0.042** | **0.058** | **0.077** | **0.103** | 0.152 |
| Man | AFMF | 0.029 | 0.036 | 0.042 | 0.048 | 0.063 | 0.074 | 0.093 | 0.119 | 0.168 |
|  | IMF | 0.012 | 0.021 | 0.030 | 0.042 | 0.054 | 0.070 | 0.081 | 0.108 | **0.137** |
|  | DAMF | 0.012 | 0.022 | 0.030 | 0.043 | 0.058 | 0.070 | 0.087 | 0.107 | 0.148 |
|  | TSF | **0.010** | 0.022 | 0.032 | 0.040 | 0.056 | 0.067 | 0.084 | 0.105 | 0.148 |
|  | ARMF | **0.010** | 0.020 | 0.029 | 0.040 | 0.053 | 0.067 | 0.084 | 0.109 | 0.144 |
|  | SFT_Lp | **0.010** | **0.018** | **0.027** | **0.037** | **0.049** | **0.062** | **0.080** | **0.104** | 0.167 |
| Peppers | AFMF | 0.015 | 0.024 | 0.028 | 0.035 | 0.044 | 0.058 | 0.077 | 0.098 | 0.156 |
|  | IMF | 0.005 | 0.010 | 0.024 | 0.027 | 0.041 | 0.053 | 0.067 | 0.090 | **0.127** |
|  | DAMF | **0.004** | 0.014 | 0.022 | 0.030 | 0.040 | 0.048 | 0.064 | 0.087 | 0.129 |
|  | TSF | 0.008 | 0.015 | 0.021 | 0.030 | 0.042 | 0.051 | 0.065 | 0.089 | 0.130 |
|  | ARMF | 0.006 | 0.013 | 0.024 | 0.028 | 0.033 | **0.045** | **0.061** | 0.089 | 0.129 |
|  | SFT_Lp | **0.004** | **0.009** | **0.016** | **0.023** | **0.029** | **0.045** | **0.061** | **0.086** | 0.144 |
| Camera | AFMF | 0.028 | 0.035 | 0.041 | 0.053 | 0.661 | 0.078 | 0.098 | 0.118 | 0.194 |
|  | IMF | 0.013 | 0.022 | 0.034 | 0.045 | 0.062 | 0.071 | 0.090 | 0.112 | 0.149 |
|  | DAMF | 0.012 | 0.023 | 0.031 | 0.051 | 0.059 | 0.070 | 0.087 | 0.108 | 0.146 |
|  | TSF | 0.012 | 0.022 | 0.032 | 0.048 | 0.059 | 0.067 | 0.085 | 0.109 | **0.138** |
|  | ARMF | 0.011 | 0.020 | 0.029 | 0.041 | 0.050 | 0.067 | 0.083 | **0.107** | 0.146 |
|  | SFT_Lp | **0.009** | **0.016** | **0.032** | **0.037** | **0.045** | **0.065** | **0.081** | 0.109 | 0.166 |
| Barbara | AFMF | 0.063 | 0.074 | 0.079 | 0.094 | 0.103 | 0.116 | 0.129 | 0.148 | 0.182 |
|  | IMF | 0.032 | 0.050 | 0.057 | 0.072 | 0.085 | 0.093 | 0.108 | 0.128 | **0.151** |
|  | DAMF | 0.034 | 0.057 | 0.071 | 0.085 | 0.099 | 0.113 | 0.129 | 0.151 | 0.182 |
|  | TSF | 0.033 | 0.049 | 0.072 | 0.085 | 0.100 | 0.118 | 0.130 | 0.143 | 0.177 |
|  | ARMF | 0.031 | 0.049 | 0.060 | 0.078 | 0.093 | 0.110 | 0.131 | 0.149 | 0.178 |

|       | SFT_Lp | **0.026** | **0.035** | **0.053** | **0.064** | **0.074** | **0.089** | **0.106** | **0.123** | 0.164 |
|-------|--------|-----------|-----------|-----------|-----------|-----------|-----------|-----------|-----------|-------|
|       | AFMF   | 0.035 | 0.037 | 0.047 | 0.056 | 0.069 | 0.087 | 0.104 | 0.126 | 0.177 |
|       | IMF    | 0.015 | 0.026 | 0.035 | 0.046 | 0.060 | 0.076 | 0.091 | **0.114** | **0.145** |
| Boat  | DAMF   | 0.014 | 0.023 | 0.038 | 0.050 | 0.062 | 0.078 | 0.098 | 0.119 | 0.153 |
|       | TSF    | 0.014 | 0.026 | 0.039 | 0.051 | 0.063 | 0.078 | 0.090 | **0.114** | 0.151 |
|       | ARMF   | 0.014 | 0.024 | **0.033** | 0.044 | 0.059 | 0.076 | 0.092 | **0.114** | 0.151 |
|       | SFT_Lp | **0.010** | **0.016** | **0.033** | **0.041** | **0.054** | **0.074** | **0.088** | 0.131 | 0.182 |

Table 5 Cost time (s) of different methods under different level noise

| Image | Method | 10% | 20% | 30% | 40% | 50% | 60% | 70% | 80% | 90% |
|-------|--------|-----|-----|-----|-----|-----|-----|-----|-----|-----|
|       | AFMF   | 12.45 | 11.68 | 11.84 | 11.42 | 11.21 | 10.91 | 10.60 | 10.85 | 12.13 |
|       | IMF    | 0.53 | 0.99 | 2.31 | 3.01 | 3.75 | 5.95 | 6.98 | 9.71 | 12.24 |
| Lena512 | DAMF | 0.25 | 0.51 | **0.70** | **0.94** | 1.18 | **1.33** | **1.60** | **1.92** | **2.38** |
|       | TSF    | **0.23** | **0.45** | 0.82 | 0.97 | **1.10** | 1.53 | 1.63 | 2.05 | 2.91 |
|       | ARMF   | 0.37 | 0.38 | 0.45 | 0.63 | 0.78 | 0.87 | 0.99 | 1.17 | 1.46 |
|       | SFT_Lp | 110.4 | 116.3 | 124.2 | 109.5 | 115.2 | 124.1 | 131.7 | 155.4 | 183.1 |
|       | AFMF   | 3.14 | 3.22 | 3.08 | 2.98 | 2.96 | 0.92 | 2.90 | 2.92 | 3.05 |
|       | IMF    | 0.28 | 0.46 | 0.94 | 1.13 | 1.76 | 2.67 | 2.90 | 3.70 | 4.71 |
| House | DAMF  | 0.10 | 0.18 | 0.21 | 0.29 | 0.36 | 0.39 | 0.47 | 0.60 | 0.81 |
|       | TSF    | **0.05** | **0.13** | **0.18** | **0.20** | **0.24** | **0.38** | **0.43** | **0.48** | **0.64** |
|       | ARMF   | 0.10 | 0.14 | 0.15 | 0.23 | 0.28 | 0.38 | 0.39 | 0.42 | 0.44 |
|       | SFT_Lp | 18.56 | 18.49 | 21.18 | 22.41 | 23.35 | 24.27 | 28.22 | 32.67 | 37.74 |
|       | AFMF   | 3.07 | 3.07 | 3.26 | 3.06 | 3.04 | 2.87 | 2.77 | 2.84 | 3.23 |
|       | IMF    | 0.24 | 0.69 | 0.98 | 1.22 | 1.59 | 1.67 | 2.55 | 3.64 | 5.89 |
| Gold hill | DAMF | 0.11 | 0.18 | 0.22 | 0.28 | 0.39 | 0.47 | 0.50 | 0.64 | 0.85 |
|       | TSF    | **0.05** | **0.11** | **0.17** | **0.26** | **0.28** | **0.40** | **0.42** | **0.51** | **0.64** |
|       | ARMF   | 0.10 | 0.12 | 0.20 | 0.22 | 0.24 | 0.32 | 0.33 | 0.48 | 0.53 |
|       | SFT_Lp | 17.93 | 21.04 | 21.36 | 22.27 | 23.26 | 24.86 | 27.63 | 30.16 | 38.03 |
|       | AFMF   | 3.27 | 3.25 | 3.17 | 3.07 | 2.98 | 2.88 | 2.86 | 2.73 | 2.96 |
|       | IMF    | 0.27 | 0.46 | 0.66 | 1.17 | 1.81 | 2.54 | 2.86 | 3.25 | 4.74 |
| Man   | DAMF   | 0.10 | 0.17 | 0.25 | 0.27 | 0.34 | 0.44 | 0.53 | 0.51 | 0.76 |
|       | TSF    | **0.06** | **0.11** | **0.19** | **0.33** | **0.31** | **0.34** | **0.45** | **0.53** | **0.60** |
|       | ARMF   | 0.10 | 0.17 | 0.20 | 0.25 | 0.27 | 0.29 | 0.34 | 0.40 | 0.44 |
|       | SFT_Lp | 18.62 | 19.28 | 20.88 | 22.14 | 24.67 | 25.44 | 27.36 | 31.16 | 39.81 |
|       | AFMF   | 3.06 | 3.20 | 3.16 | 2.90 | 3.06 | 3.01 | 2.73 | 2.97 | 3.23 |
|       | IMF    | 0.25 | 0.47 | 0.85 | 1.28 | 1.88 | 2.72 | 2.54 | 4.09 | 4.69 |
| Peppers | DAMF | 0.11 | 0.21 | 0.21 | 0.29 | 0.33 | 0.38 | 0.57 | 0.60 | 0.73 |
|       | TSF    | **0.05** | **0.12** | **0.16** | **0.21** | **0.35** | **0.40** | **0.38** | **0.50** | **0.61** |
|       | ARMF   | 0.10 | 0.16 | 0.19 | 0.24 | 0.33 | 0.32 | 0.40 | 0.47 | 0.50 |
|       | SFT_Lp | 17.96 | 20.19 | 21.02 | 23.30 | 24.85 | 25.27 | 28.16 | 33.64 | 40.09 |
|       | AFMF   | 4.03 | 3.90 | 3.82 | 3.83 | 3.53 | 3.26 | 3.43 | 3.44 | 3.98 |
|       | IMF    | 0.25 | 0.47 | 0.79 | 1.10 | 1.37 | 1.68 | 2.41 | 3.56 | 4.12 |
| Camera | DAMF  | 0.10 | 0.16 | 0.21 | 0.25 | 0.33 | 0.36 | 0.49 | 0.59 | 0.68 |
|       | TSF    | **0.05** | **0.08** | **0.15** | **0.20** | **0.31** | **0.29** | **0.39** | **0.45** | **0.59** |
|       | ARMF   | 0.10 | 0.19 | 0.26 | 0.23 | 0.27 | 0.24 | 0.37 | 0.46 | 0.47 |

|  | SFT_Lp | 18.75 | 20.22 | 22.08 | 22.79 | 23.77 | 26.39 | 30.17 | 32.28 | 40.05 |
|---|---|---|---|---|---|---|---|---|---|---|
| Barbara | AFMF | 3.65 | 3.59 | 3.68 | 3.78 | 3.65 | 3.17 | 3.39 | 3.20 | 3.77 |
|  | IMF | 0.25 | 0.40 | 0.64 | 1.14 | 1.74 | 1.58 | 2.47 | 3.27 | 4.67 |
|  | DAMF | 0.09 | 0.18 | 0.23 | 0.33 | 0.35 | 0.49 | 0.57 | 0.60 | 0.83 |
|  | TSF | **0.04** | **0.11** | **0.14** | **0.18** | **0.22** | **0.32** | **0.34** | **0.41** | **0.49** |
|  | ARMF | 0.12 | 0.14 | 0.19 | 0.23 | 0.26 | 0.30 | 0.40 | 0.36 | **0.49** |
|  | SFT_Lp | 18.39 | 20.01 | 21.82 | 22.62 | 24.53 | 25.56 | 26.68 | 33.59 | 41.06 |
| Boat | AFMF | 3.64 | 4.20 | 3.97 | 3.65 | 3.67 | 3.46 | 3.15 | 3.27 | 3.81 |
|  | IMF | 0.25 | 0.44 | 0.84 | 1.18 | 1.52 | 1.61 | 2.34 | 2.75 | 4.85 |
|  | DAMF | 0.09 | 0.18 | 0.25 | 0.30 | 0.32 | 0.53 | 0.54 | 0.58 | 0.77 |
|  | TSF | **0.05** | **0.12** | **0.14** | **0.20** | **0.26** | **0.28** | **0.34** | **0.43** | 0.54 |
|  | ARMF | 0.09 | 0.15 | 0.16 | 0.21 | **0.26** | 0.29 | **0.34** | 0.47 | **0.48** |
|  | SFT_Lp | 18.55 | 21.00 | 21.59 | 23.04 | 25.42 | 26.54 | 27.94 | 32.93 | 40.77 |

## 4.4 Ablation experiments

To further evaluate the critical parts of the proposed model, we performed ablation experiments for the MCA framework, mask matrix, sparse transform, and $l_p$ quasi-norm.

### 4.4.1 Ablation of the MCA

We compared the proposed model with the SR-based denoising model (the objective function is $F = \underset{F}{\mathrm{argmin}}\, \alpha_0 \|M \circ F - G\|_{p_0}^{p_0} + \alpha_1 \|D(F)\|_{p_1}^{p_1}$), in which the recovery image is not considered as the sum of the cartoon and texture parts. Although the reconstructed qualities of the cartoon part shown in Figure 8 (c) and the texture part shown in Figure 8 (d) are less than those of the SR-based model, the final reconstructed image is better than the recovery image of the SR-based model.

It is difficult to find out the difference between Figure 8 (b) and Figure 8 (e) visually. Thus we evaluate the performance by the evaluation indexes. The PSNR obtained by the proposed model is 0.5 dB higher than that obtained by the SR-based model. The SSIM obtained by the proposed model is as the same as that obtained by the SR-based model. The GMSD obtained by the proposed model is smaller than that obtained by the SR-based model. Thus, according to the indexes used in this paper, the MCA-based model outperforms the SR-based model. It is because the MCA framework restores the image hierarchically. In this way, the low-frequency component (in the cartoon part) and the high-frequency component (in the texture part) do not interfere with obtaining a more detailed restored image.

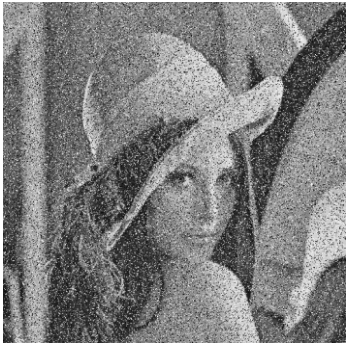 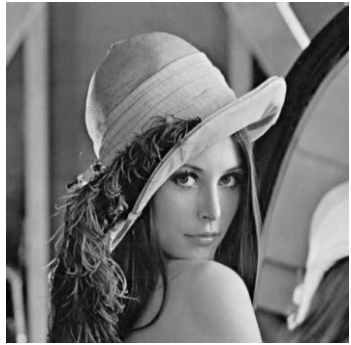 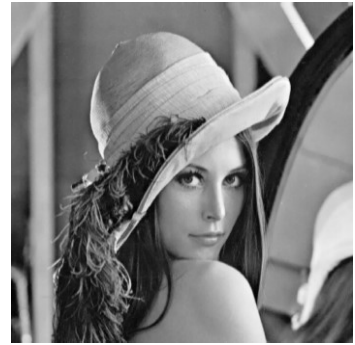

(a)　　　　　　　　　　　　(b)　　　　　　　　　　　　(c)

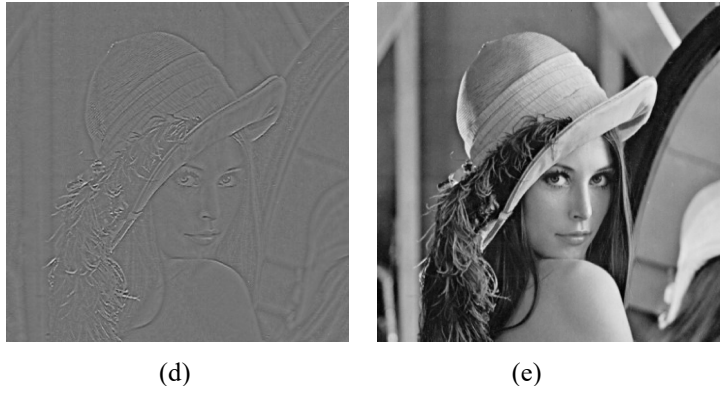

(d)            (e)

Figure 8 Results of the ablation experiments. (a) "Lena512" under 20% salt and pepper noise (PSNR=12.44, SSIM=0.269, and GMSD=0.295); (b) reconstructed image (PSNR=41.68, SSIM=**0.996**, and GMSD= 0.006) by the objective function, $\boldsymbol{F}=\underset{\boldsymbol{F}}{argmin}\ \alpha_0 \|\boldsymbol{M} \circ \boldsymbol{F} - \boldsymbol{G}\|_{p_0}^{p_0} + \alpha_1 \|\boldsymbol{D}(\boldsymbol{F})\|_{p_1}^{p_1}$ ; (c) reconstructed cartoon part (PSNR=38.38, SSIM=0.989, and GMSD=0.013) by the proposed model; (d) reconstructed texture part (PSNR=5.67, SSIM= $1.82 \times 10^{-4}$, and GMSD=0.319) by the proposed model; (e) reconstructed image (PSNR=**42.18**, SSIM=**0.996**, and GMSD=**0.005**) by the proposed model.

**4.4.2 Ablation of the mask matrix**

This section describes the mask matrix's effects in the proposed model. Based on the test image shown in Figure 5 (a), Figure 9 shows the result obtained by the objective function without the mask matrix (the objective function is $[\boldsymbol{F}_T,\boldsymbol{F}_C]=\underset{[\boldsymbol{F}_T,\boldsymbol{F}_C]}{argmin}\ \alpha_0 \|\boldsymbol{F}_T+\boldsymbol{F}_C-\boldsymbol{G}\|_{p_0}^{p_0} + \alpha_1 \|\boldsymbol{D}(\boldsymbol{F}_C)\|_{p_1}^{p_1} + \alpha_2 \|\boldsymbol{D}(\boldsymbol{F}_T)\|_{p_2}^{p_2}$ ) and the result obtained by the proposed model.

As shown in Figure 9, the quality of the reconstructed model without the mask matrix is considerably reduced. The main reason for this phenomenon is that the mask matrix contains the noise position information, which plays a role in protecting the area not polluted by noise. In this way, the clean data are protected. In addition, the area contaminated by noise can be sparsely represented by the Framelet coefficients. Thus, the mask matrix is significant for preserving the image details.

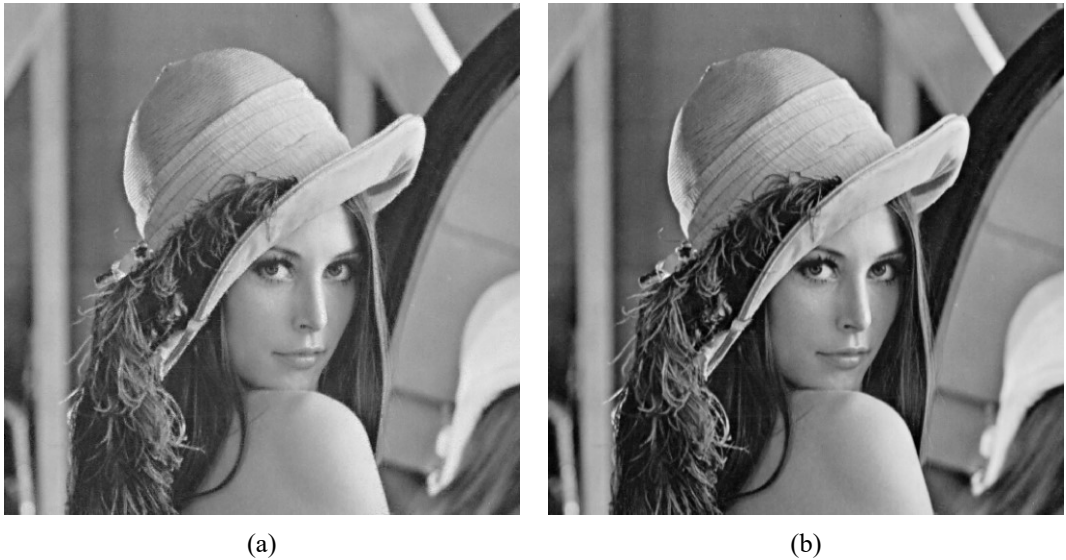

(a)            (b)

Figure 9 Results of the ablation experiments. (a) Reconstructed image (PSNR= 34.90, SSIM=0.987, and GMSD=0.019) by the objective function, $[\boldsymbol{F}_T,\boldsymbol{F}_C]=\underset{[\boldsymbol{F}_T,\boldsymbol{F}_C]}{argmin}\ \alpha_0 \|\boldsymbol{F}_T+\boldsymbol{F}_C-\boldsymbol{G}\|_{p_0}^{p_0} + \alpha_1 \|\boldsymbol{D}(\boldsymbol{F}_C)\|_{p_1}^{p_1} + \alpha_2 \|\boldsymbol{D}(\boldsymbol{F}_T)\|_{p_2}^{p_2}$ ;

(b) reconstructed image (PSNR= **42.18**, SSIM=**0.996**, and GMSD= **0.005**) by the proposed model.

### 4.4.3 Ablation of the sparse transform

We compared the wavelet-based model (the objective function is $[\boldsymbol{F}_T, \boldsymbol{F}_C] = \underset{[\boldsymbol{F}_T, \boldsymbol{F}_C]}{\mathrm{argmin}}\, \alpha_0 \|\boldsymbol{M} \circ \boldsymbol{F}\ \boldsymbol{F}\ -\boldsymbol{G}\|_{p_0}^{p_0} + \alpha_1 \|\boldsymbol{W}(\boldsymbol{F}_C)\|_{p_1}^{p_1} + \alpha_2 \|\boldsymbol{W}(\boldsymbol{F}_T)\|_{p_2}^{p_2}$) and the proposed model. The wavelet basis is the Haar wavelet [60]. The WT and inverse WT were implemented using the Mallat algorithm [61]. The results of the ablation experiment are shown in Figure 10.

As shown in Figure 10, the image quality restored using the Haar WT is not as high as that recovered by the SFT. Because the SFT does not have the down-sample operation in the WT, the Framelet coefficients do not need to be up-sampled by padding zeros when reconstructing the image; this avoids the image information loss in the WT and inverse WT. Compared with the WT, SFT adds a high-pass filter and can decompose the image texture information more accurately than the WT. Therefore, the quality of the recovery image obtained via SFT was higher than that of the WT.

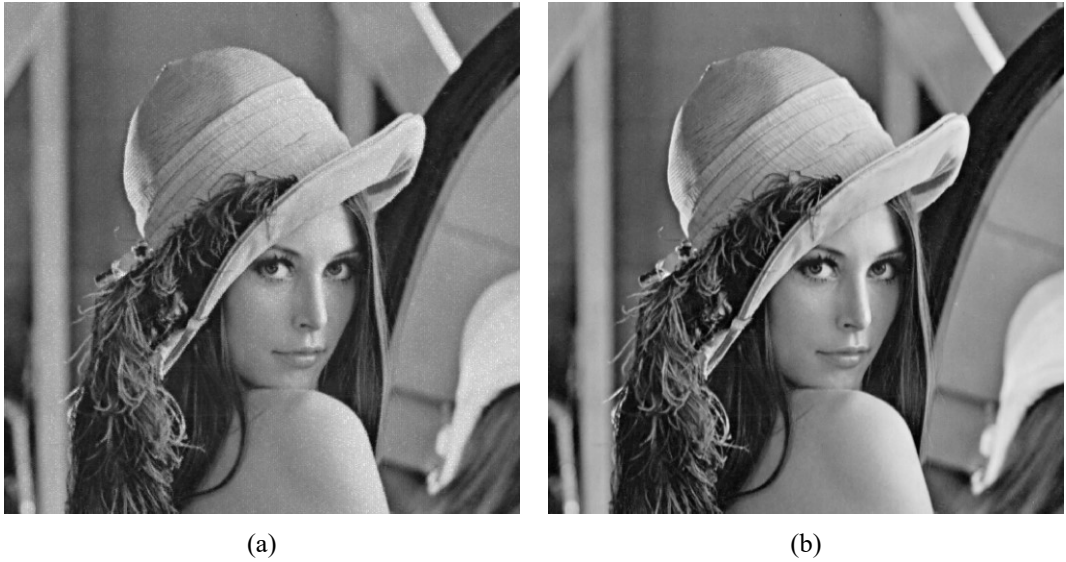

(a) (b)

Figure 10 Results of the ablation experiments. (a) Reconstructed image (PSNR=33.14, SSIM=0.962, and GMSD=0.028) by the objective function, $[\boldsymbol{F}_T, \boldsymbol{F}_C] = \underset{[\boldsymbol{F}_T, \boldsymbol{F}_C]}{\mathrm{argmin}}\, \alpha_0 \|\boldsymbol{M} \circ \boldsymbol{F}\ \boldsymbol{F}\ -\boldsymbol{G}\|_{p_0}^{p_0} + \alpha_1 \|\boldsymbol{W}(\boldsymbol{F}_C)\|_{p_1}^{p_1} + \alpha_2 \|\boldsymbol{W}(\boldsymbol{F}_T)\|_{p_2}^{p_2}$; (b) reconstructed image (PSNR=**42.18**, SSIM=**0.996**, and GMSD=**0.005**) by the proposed model.

### 4.4.4 Ablation of the $l_p$ quasi-norm

We compared the proposed model with a model using the $l_1$ norm (the objective function is $[\boldsymbol{F}_T, \boldsymbol{F}_C] = \underset{[\boldsymbol{F}_T, \boldsymbol{F}_C]}{\mathrm{argmin}}\, \alpha_0 \|\boldsymbol{M} \circ \boldsymbol{F}\ \boldsymbol{F}\ -\boldsymbol{G}\|_1 + \alpha_1 \|\boldsymbol{D}(\boldsymbol{F}_C)\|_1 + \alpha_2 \|\boldsymbol{D}(\boldsymbol{F}_T)\|_1$). The results of the ablation experiment are shown in Figure 11. The PSNR obtained by the proposed model was 0.58 dB higher than that obtained by the model with the $l_1$ norm while the SSIM and the GMSD obtained by the proposed method are the same as the one of the compared model.

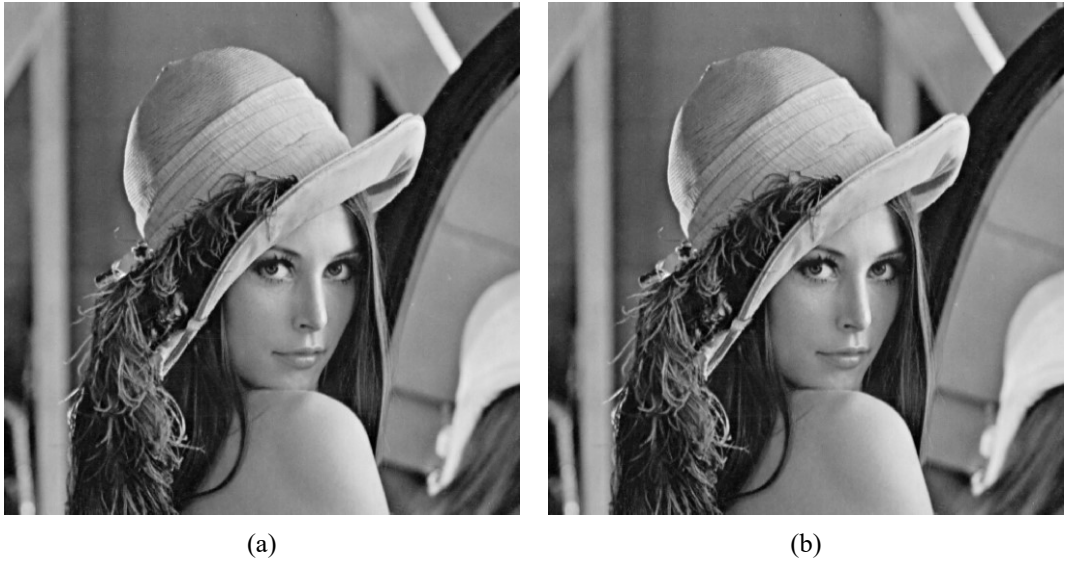

(a)                  (b)

Figure 11 Results of the ablation experiments. (a) Reconstructed image (PSNR=41.60, SSIM=**0.996**, and GMSD=**0.005**) by the objective function, $[\boldsymbol{F}_T, \boldsymbol{F}_C] = \underset{[\boldsymbol{F}_T, \boldsymbol{F}_C]}{argmin}\ \alpha_0 \|\boldsymbol{M} \circ \boldsymbol{F}\ \boldsymbol{F} - \boldsymbol{G}\|_1 + \alpha_1 \|\boldsymbol{D}(\boldsymbol{F}_C)\|_1 + \alpha_2 \|\boldsymbol{D}(\boldsymbol{F}_T)\|_1$; (b) reconstructed image (PSNR=**42.18**, SSIM=**0.996**, and GMSD=**0.005**) by the proposed model.

Because the $l_p$ quasi-norm can express the sparsity more accurately than the $l_1$ norm, the model with the $l_p$ quasi-norm achieves a better-reconstructed result. Moreover, the $l_p$ quasi-norm has more flexible degrees of freedom and can be adjusted with the change in noise level to obtain the best image restoration effect. Therefore, the proposed model has a better image restoration ability.

## 5. Conclusion

In this study, we propose a salt-and-pepper noise removal method based on MCA with the $l_p$ quasi-norm. In the proposed method, we first explore the noise position information by capturing the amplitude characteristics of salt and pepper noise to determine the mask matrix. The image is decomposed into cartoon and texture parts by introducing the mask matrix and the SFT into the MCA framework. The $l_p$ quasi-norm is adopted to depict the sparsity of salt and pepper noise and the Framelet coefficients. Several experiments were conducted to verify the effectiveness of the proposed model. The main findings of this study are summarized as follows:

(1) Under different noise levels in the test image, the PSNR, SSIM, and GMSD of the recovered images obtained by employing the proposed model were better than those of other denoising methods, demonstrating that the denoising effect of the proposed model is excellent.

(2) The dynamic iteration curves showed that the proposed method could obtain a convergent solution in relatively fewer steps.

(3) Ablation experiments showed that all the key components (MCA framework, mask matrix, SFT, and $l_p$ quasi-norm) of the proposed model significantly contribute to the image denoising performance.

However, the proposed model has some limitations: 1) the parameter selection is adjusted manually, and finding the optimal for the best denoising performance is challenging; 2) the computational efficiency of the model is low. Thus, we will focus on solving these limitations in further study.

**Author Contributions**

Yingpin Chen proposed the idea of this study, wrote the original draft, and organized the team of the authors. Yuming Huang implemented the proposed method by programing. Lingzhi Wang, Huiying

Huang, and Jianhua Song were responsible for the experiments. Chaoqun Yu and Yanping Xu were responsible for language enhancement work.

**Appendix A**

The corresponding objective function of $F_C$ is as follows:

$$J_{F_C} = \min_{F_C}\{-\langle \lambda_0 \tilde{Q} \quad Q \quad \cdot [M \circ F \quad F \quad G \quad \hat{Q} \quad M \circ F \quad F \quad G \\ -\langle \lambda_1 \tilde{Q} \quad Q \quad \cdot D(F_C)\rangle + \frac{\lambda_1}{2}\|Q_1^{(k)} - D(F_C)\|_2^2\}. \tag{A.1}$$

The method of completing the square can be used to transform equation (A.1) to

$$F_C = \arg\min_{F_C}\{\frac{\lambda_0}{2}\|Q_0^{(k)} - [M \circ F \quad F_) - G] - \tilde{Q} \quad Q \quad D F \quad \tilde{Q} \tag{A.2}$$

Let $\dfrac{\partial J_{F_C}}{\partial F_C} = 0$; we have

$$\lambda_0 M \circ \quad M \circ F \quad F) - G] + \tilde{Q} \quad Q \quad D \quad D F \quad \tilde{Q} \quad Q \tag{A.3}$$

where $D^{-1}$ represents the inverse SFT operator.

We organize equation (A.3) as follows:

$$\lambda_0 M \circ M \circ F \quad F \quad M \circ G \quad M \circ Q \quad Q \quad D Q \quad Q \quad M \circ M \circ F \tag{A.4}$$

As $M \circ M \circ F \quad M \circ F$, we have

$$\lambda_0 M \circ F \quad F \quad M \circ G \quad M \circ Q \quad Q \quad D Q \quad Q \quad M \circ F \tag{A.5}$$

**Appendix B**

The corresponding objective function of $F_T$ is as follows:

$$J_{F_T} = \min_{F_T}\{-\langle \lambda_0 \tilde{Q} \quad Q \quad \cdot [M \circ F \quad F^{+1)}) - G]\rangle \\ + \frac{\lambda_0}{2}\|Q_0^{(k)} - [M \circ F \quad F \quad - G]\|_2^2 - \langle \lambda_2 \tilde{Q} \quad Q \quad \cdot D(F_T)\rangle + \frac{\lambda_2}{2}\|Q_2^{(k)} - D(F_T)\|_2^2\}. \tag{B.1}$$

The method of completing the square can be used to transform equation (B.1) to

$$J_{F_T} = \min_{F_T}\left\{\frac{\lambda_0}{2}\|Q_0 - [M \circ F \quad F \quad - G] - \tilde{Q} \quad Q \quad D F \quad \tilde{Q}\right\} \tag{B.2}$$

Let $\dfrac{\partial J_{F_T}}{\partial F_T} = 0$; then, we have

$$\lambda_0 M \circ F \quad F \quad M \circ G \quad M \circ Q \quad Q \quad D Q \quad Q \quad M \circ F \tag{B.3}$$

**Appendix C**

The $Q_0$ sub-problem is as follows:

$$J_{Q_0} = \min_{Q_0}\left\{\alpha_0\|Q_0\|_{p_0}^{p_0} - \langle \lambda_0 \tilde{Q} \quad Q \quad M \circ F \quad F \quad G \quad \hat{Q} \quad M \circ F \quad F \quad G\right. \tag{C.1}$$

The method of completing the square can be used to transform equation (C.1) to

$$J_{Q_0} = \min_{Q_0} \{\alpha_0 \|Q_0\|_{p_0}^{p_0} - 2\left(\frac{\lambda_0}{2}\right)\langle \tilde{Q} \quad Q \quad M \circ F \quad F_C^{(k+1)}) - G]\rangle$$

$$+ \frac{\lambda_0}{2}\|Q_0 - [M \circ F \quad F_C^{(k+1)}) - G]\|_2^2 + \frac{\lambda_0}{2}\|\tilde{Q} \quad _{\|2} \quad _2 \quad \tilde{Q}\quad_{\|2}$$

$$= \min_{Q_0} \{\alpha_0 \|Q_0\|_{p_0}^{p_0} - \frac{\lambda_0}{2}\|\tilde{Q}\quad_{\|2} \quad \quad \quad \quad \quad \quad \quad \quad \quad \quad \quad \quad \text{(C.2)}$$

$$+ \frac{\lambda_0}{2}\{\|Q_0 - [M \circ F \quad F \quad G \quad \langle Q \quad Q \quad M \circ F \quad ^{+1)} + F_C^{(k+1)}) - G]\rangle + \|\tilde{Q}\quad_{\|2}$$

$$= \min_{Q_0} \{\alpha_0 \|Q_0\|_{p_0}^{p_0} + \frac{\lambda_0}{2}\|Q_0 - [M \circ F \quad F_C^{(k+1)}) - G] - \tilde{Q}\quad_{\|2} \quad _2 \quad \tilde{Q}\quad_{\|2}$$

As all the variables are decoupled, the objective function of $Q_0$ is equal to

$$J_{Q_0} = \min_{Q_0}\left\{\alpha_0 \|Q_0\|_{p_0}^{p_0} + \frac{\lambda_0}{2}\|Q_0 - [M \circ F \quad F_C^{(k+1)}) - G] - \tilde{Q}\quad_{\|2}\right\} \quad \text{(C.3)}$$